\documentclass{aa}  

\newcommand{\sii}{[\ion{S}{ii}]}
\newcommand{\ha}{H$\alpha$}

\usepackage{natbib,twoopt}
\usepackage{gensymb}
\usepackage[super]{nth}
\usepackage{txfonts}
\usepackage{graphicx}
\usepackage[labelfont=bf]{caption}
\usepackage{subcaption}
\usepackage{changepage}
\usepackage{float}
\usepackage{orcidlink}
\usepackage{stfloats}
\usepackage{booktabs} 
\usepackage{threeparttable}
\usepackage{hyperref} 
\hypersetup{
    breaklinks=true,
    colorlinks,
    citecolor=blue,
    filecolor=black,
    linkcolor=blue,
    urlcolor=black
}
\bibpunct{(}{)}{;}{a}{}{,}             
\makeatletter
  \newcommandtwoopt{\citeads}[3][][]{\href{http://adsabs.harvard.edu/abs/#3}%
    {\def\hyper@linkstart##1##2{}%
     \let\hyper@linkend\@empty\citealp[#1][#2]{#3}}}
  \newcommandtwoopt{\citepads}[3][][]{\href{http://adsabs.harvard.edu/abs/#3}%
    {\def\hyper@linkstart##1##2{}%
     \let\hyper@linkend\@empty\citep[#1][#2]{#3}}}
  \newcommandtwoopt{\citetads}[3][][]{\href{http://adsabs.harvard.edu/abs/#3}%
    {\def\hyper@linkstart##1##2{}%
     \let\hyper@linkend\@empty\citet[#1][#2]{#3}}}
  \newcommandtwoopt{\citeyearads}[3][][]%
    {\href{http://adsabs.harvard.edu/abs/#3}
    {\def\hyper@linkstart##1##2{}%
     \let\hyper@linkend\@empty\citeyear[#1][#2]{#3}}}
\makeatother

\newcommand{\spbub}{DEM L36}

\begin{document} 
   \title{The {\it XMM-Newton} view of the \spbub{} superbubble in the Large Magellanic Cloud\thanks{Based on observations obtained with {\it XMM-Newton}, an ESA science mission with instruments and contributions directly funded by ESA Member States and NASA.}}

   \author{Pawel M. Janas\inst{1}\corrauth{Pawel.Janas@mu.ie} \orcidlink{0009-0002-8388-6216}
        \and
        Patrick J. Kavanagh \inst{1}\email{Patrick.Kavanagh@mu.ie} \orcidlink{0000-0001-6872-2358}
        \and
        Manami Sasaki \inst{2}\email{manami.sasaki@fau.de} \orcidlink{0000-0001-5302-1866}
        \and
        Miroslav D. Filipovi\'c \inst{3}\email{M.Filipovic@westernsydney.edu.au} \orcidlink{000-0002-4990-9288}
        \and
        Chandreyee Maitra \inst{4,5}\email{cmaitra@mpe.mpg.de} \orcidlink{0000-0002-0766-7313}
        \and
        Frank Haberl \inst{4}\email{fwh@mpe.mpg.de} \orcidlink{0000-0002-0107-5237}
        \and
        Pierre Maggi \inst{6}\email{pierre.maggi@astro.unistra.fr} \orcidlink{0000-0001-5612-5185}
        \and
        Sean Points \inst{7}\email{sean.points@noirlab.edu} \orcidlink{0000-0002-4596-1337}
        \and
        Zachary J. Smeaton \inst{3}\email{19594271@student.westernsydney.edu.au}  \orcidlink{0009-0009-7061-0553}}
   \institute{Department of Physics, Maynooth University,
              Maynooth, Co. Kildare, Ireland
              \and
              Dr. Karl Remeis Observatory, Erlangen Centre for Astroparticle Physics, Friedrich-Alexander-Universit\"{a}t Erlangen-N\"{u}rnberg, Sternwartstrasse 7, 96049 Bamberg, Germany
              \and
              Western Sydney University, Locked Bag 1797, Penrith South DC, NSW 2751, Australia
              \and
              Max Planck Institute for Extraterrestrial Physics, Giessenbachstrasse 1, 85748 Garching, Germany
              \and
              Inter University Centre for Astronomy \& Astrophysics, Ganeshkhind, Pune 411007, India
              \and
              Observatoire Astronomique de Strasbourg, Université de Strasbourg, CNRS, 11 rue de l’Universite , 67000 Strasbourg, France
              \and
              National Science Foundation National Optical-Infrared Astronomy Research Laboratory: La Serena, Chile
              }

   \date{Received --/ Accepted --}

\abstract
{Massive star-forming regions are important laboratories for our understanding of stellar feedback mechanisms in galaxies. They often contain superbubbles, large cavities of hot, shock-heated gas powered by massive stellar winds and subsequent supernovae. 
}
{We present the first multi-wavelength analysis of the \spbub{} superbubble in the Large Magellanic Cloud. We aim to characterise its physical properties for comparison to model predictions. Additionally, we compare the \spbub{} superbubble with other Galactic and Magellanic Cloud superbubbles.}
{We use data from \emph{XMM-Newton}, supplemented by data from the Magellanic Cloud Emission Line Survey, the Dark Energy Camera Magellanic Cloud Emission Line Survey, the \emph{Spitzer} Space Telescope, and the Australia Telescope Compact Array. We perform a multi-wavelength morphological analysis to understand the shape and structure of the superbubble and X-ray spectral analysis to derive physical parameters of the hot gas. }
{We find significant X-ray emission in \spbub{}, dominated by a central bubble region mostly confined by a thin H$\alpha$ filamentary shell. This has two distinct blowout regions to the east and west, with a possible smaller blowout in the south. Spectral analysis shows the X-ray emission is well fit with a thermal plasma model in collisional ionisation equilibrium with a temperature of $kT=0.17~(0.16-0.18)$ keV and unabsorbed luminosity $L_{X~(0.5-7.0~\mathrm{keV})}=3.62\times 10^{35}$~erg~s$^{-1}$. Unlike other LMC superbubbles, we find no clear powering cluster at the centre, though relatively small clusters near the edge are likely supplying the mechanical input at the present epoch. We use photometric data of these clusters to estimate cluster masses, ages, and mechanical inputs to compare with the energy stored in the superbubble. We find that these clusters have supplied more energy over their lifetime than is currently observed in the superbubble, indicative of the so-called growth rate discrepancy problem.}
{It is likely the \spbub{} superbubble was initially shaped by a central massive stellar population that has since dispersed and largely lost to supernovae, with smaller clusters at the edges of the swept-up shell being currently supplying the energy input.}


   \titlerunning{{\it XMM-Newton} observations of the DEM L36 SB in the LMC}
   \authorrunning{P. M. Janas et al.}
   \maketitle
   \nolinenumbers
%

\section{Introduction}
The evolution of the interstellar medium (ISM) is dominated by feedback from massive stars and later by supernovae and their remnants, regulating star formation efficiency, driving turbulence, and triggering or suppressing subsequent star formation on multiple scales. Winds and SNRs of OB stars in massive stellar clusters drive the evolution of massive star forming regions (MSFRs) and often form `superbubbles' (SBs) - huge cavity regions within the ISM with diameters of 100-1000 pc which are filled with a hot, shock-heated X-ray emitting gas with temperatures of $\sim10^6$ K \citep[see][for a review]{kavanagh2020review}. This hot gas is surrounded by a `supershell' of swept-up ISM which is bright in optical emission, primarily powered by photoionisation from the intense UV radiation from the massive stars. These objects were initially described by the analytical models of \cite{maclow1988superbubbles} and \cite{chu1990model}, extensions of the wind-blown bubble model of \citet{weaver1977interstellar}, which assumed a continuous stellar input and uniform ambient interstellar medium (ISM).

However, observations of SBs and constraints on the mechanical input by their powering stellar populations deviate significantly from model predictions with the observed SBs being smaller than expected \citep[e.g.][]{Oey1996}. This is known as the `growth rate discrepancy'. Various resolutions to this issue have been proposed such as a correction to ambient pressure \citep{Oey2004}, blowouts \citep[e.g.][]{Breit2006}, radiative cooling \citep[e.g.][]{Kim2017}, and losses to cosmic ray acceleration \citep{Butt2008}. Another observed deviation from the models is that X-ray bright superbubbles have higher supershell expansion velocities than predicted \citep{Oey1996}, which likely results from the time variability of the mechanical input due to interior supernovae \citep{Oey2004}. To further our understanding of SB evolution and the contributions to the growth rate discrepancy, the identification and analysis of new objects is essential to provide empirical constraints on both analytical and numerical models. This not only broadens statistics on SBs and their properties but also reveals the diversity of feedback outcomes.

As the interior of SBs are dominated by soft X-ray emission in the 0.1 - 2.0 keV band, Galactic objects are often obscured due to large foreground absorbing columns \citep{kavanagh2020review}. Nearby galaxies, and the Large Magellanic Cloud (LMC) in particular, are excellent alternatives for the study of massive star forming regions and SBs. The LMC is a nearby irregular dwarf galaxy exhibiting spiral features at a distance of $49.6\pm 0.63$ kpc \citep{pietrzynski2019distance} with a low inclination angle $i=30\degree.7$ \citep{nikolaev2004geometry}. The LMC, which has a low metallicity environment and boasts a low foreground extinction value (average Galactic foreground $N_H=0.7\times 10^{21}$ cm$^{-2}$), makes it the ideal laboratory for observing SBs across the wavelength regimes. These viewing conditions enable the analysis of diffuse X-ray emitting gas and to constrain the mechanical output from the massive stellar population. There are several known X-ray SBs in the LMC, e.g. \citep{cooper2004energy,jaskot2011observational,sasaki2011xmm,kavanagh2012xmm}.

In this paper, we outline the discovery of a new X-ray SB located in the \spbub{} \citep{breysacher_1999_catalogue, henize_1956_deml36} \ion{H}{II} region with RA (J2000) = 4h 56m 45.47s, Dec = -69$\degree$ 28' 44.38", on the cusp of the LMC-7 supergiant shell \citep{meaburn1980shells}. Interestingly, this SB has no clear young massive stellar cluster as a powering source. We analyse possible powering stellar populations around the region using photometric data from \cite{zaritsky2004magellanic}. We present an analysis using X-ray data from \emph{XMM-Newton}, optical data from the Magellanic Cloud Emission Line Survey (MCELS), infrared data from \emph{Spitzer} and 21 cm radio data from the Australia Telescope Compact Array (ATCA). These data enable us to gain insights into the physical characteristics and processes of the SB, assess the magnitude of the growth rate discrepancy problem for this object, and allow us to compare it to other known SBs.

In Sect. \ref{sec:sec2} we outline our data and the reduction tasks undertaken. In Sect. \ref{sec:sec3} we describe the X-ray spectral analysis of the superbubble region. In Sect. \ref{sec:sec4}, we discuss the multi-wavelength morphology of the complex, interpret the results from spectral fitting, make comparisons to other superbubbles in both our Galaxy and the LMC, and compute the energy budget of the region by examining the H$\alpha$ supershell and the stellar content inside the superbubble. Finally, the summary of the work is presented in Sect. \ref{sec:sec5}.

\section{Observations and data reduction}
\label{sec:sec2}
\subsection{X-ray data}
The \spbub{} region has been observed twice with \emph{XMM-Newton} on the \nth{11} April 2020 and \nth{6} November 2017 (proposal IDs: 0841660301, PI Kavanagh; 0804550101, PI Maggi) with total integration times $\sim$55 ks and $\sim$34 ks, respectively. These data have previously been used to study the Be X-ray binary (BeXRB) pulsar J0457 in \cite{haberl2022binaries}, which is located at the eastern edge of \spbub{}. All observations were made using the Medium filter in Full Window mode. We reduced the data using the Science Analysis System (SAS) from ESA\footnote{https://www.cosmos.esa.int/web/xmm-newton/sas} version 21.0.0 with the latest Current Calibration Files (CCFs), and the standard recipe of tasks provided by SAS. This included creating concatenated event lists from the observation data files (ODFs) for both the European Photon Imaging Camera (EPIC)-pn and EPIC-MOS detectors, removing the flaring particle background in the 10 - 12 keV band by creating Good Time Interval (GTI) files using \texttt{tabgtigen}, and performing source detection using \texttt{edetect\_chain} to identify and mask any point sources visible in the diffuse region of the \spbub{} SB. Our source detection was run over five energy bands: 0.2 - 0.5 keV, 0.5 - 1.0 keV, 1.0 - 2.0 keV, 2.0 - 4.5 keV and 4.5 - 12.0 keV. The exposure times after filtering for both observations are listed in Table \ref{tab:exptimes}. Given the reduced coverage of the diffuse X-ray emission associated with \spbub{} by EPIC-MOS1 damaged CCDs, we only used data from EPIC-pn and EPIC-MOS2

\begin{table}[h]
    \centering
    \caption{Observation information and flare-filtered exposure times for the datasets used in our analysis.}
    \begin{tabular}{rrrc}
    \toprule
        Obs. ID & EPIC & Start Date & Effective Exposure (ks)\\ \midrule
        0841660301 & pn & 04-2020 & 21.4 \\
        0841660301 & MOS2 &  & 36.2 \\
        0804550101 & pn & 11-2017 & 28.0 \\
        0804550101 & MOS2 & & 33.7 \\
    \bottomrule
    \end{tabular}
    \label{tab:exptimes}
\end{table}

\subsubsection{Imaging}
\label{sec:particle-bkg}
To remove the contribution of the quiescent particle background (QPB) from our science images, we used the SAS tasks \texttt{evqpb} and \texttt{evselect}. The task \texttt{evqpb} uses the attitude history file and ODF information to gather Filter Wheel Closed (FWC) data for the given observation and oversamples it using a given oversampling factor. This is done because the associated event list created by \texttt{evqpb} contains fewer photons than the observation science data. For our data, this sampling factor was set to 2. We subtracted these images from the corresponding science exposures.

We created smoothing templates for each image using the full energy band (0.2 - 12 keV) images. We set a desired signal-to-noise ratio (S/N) of 5 with minimum/maximum widths of 4 $\sigma$ and 20 $\sigma$ respectively for the normalised Gaussian convolvers. These were used to smooth the images in each energy band using the exposure maps as weight images. The inclusion of these weight images prevents large noise fluctuations in areas of low exposure and eliminates a `drooping' effect at the edge of the images. Our smoothed images were converted to count rate images by dividing by the exposure maps. 

\begin{figure*}[!ht]
    \centering
    \includegraphics[width=\textwidth]{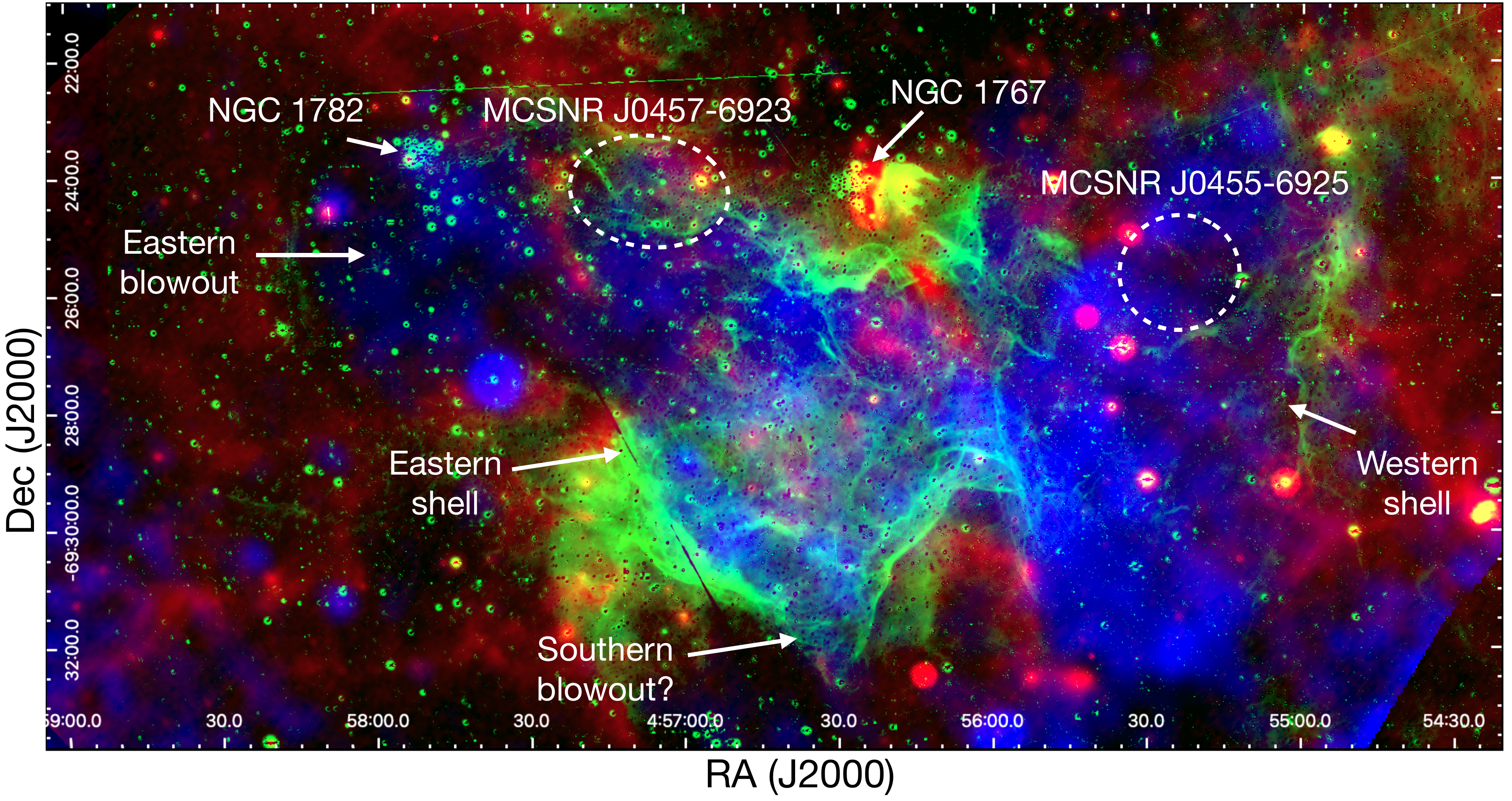}
    \caption{Three colour image of the \spbub{} superbubble. Red corresponds to a MIPS image at 24 $\mu$m, green to the DeMCELS continuum subtracted H$\alpha$ image, and blue to a 0.3 -- 1.0 keV EPIC mosaic. Annotated are two distinct blowout regions seen to the east and west of the superbubble where the H$\alpha$ shell is broken. An emerging blowout region or break in the optical shell is also seen to the south of the central bubble region. The dark line visible in the H$\alpha$ image near the eastern shell is the residual of a likely satellite trail during observations subtracted out during continuum subtraction. The candidate supernova remnants (SNRs) MCSNR~0457-6923 \citep{bozzetto2017statistical} and MCSNR~0455-6925 \citep{Cotton2026} are indicated by the white dashed ellipses, see Sect.~\ref{sect:mcsnr_cand} for details.}
    \label{fig:multi-wavelength}
\end{figure*}

Once all science exposures were filtered for background flares, QPB subtracted, and smoothed, we created mosaics of EPIC-pn and EPIC-MOS2 images by combining data from both observations using the SAS task \texttt{emosaic}. The combined 0.3-1.0 keV image is shown in Fig. \ref{fig:multi-wavelength}.

\subsubsection{Spectral extraction}
The X-ray mosaic constructed from the combined observation data (Fig. \ref{fig:multi-wavelength}) showed a large diffuse region with a complex morphology. Comparison with multi-wavelength data (Sect.~\ref{sec:sb-morphology}) suggests that the X-ray emission traces the SB region in the east, as well as a large shell in the west. We defined extraction regions for each using a contour map with the minimum being set to 3$\sigma$ above the mean background level which was taken from adjacent regions free of detected sources. We used the aperture photometry utility from the package \texttt{photutils} \citep{larry_bradley_2024_10967176} to determine background statistics. The western and eastern shells were located on the edges of the FOVs of ObsID~0804550101 and ObsID~0841660301, respectively, providing only partial coverage. For this reason, we only consider spectra extracted from ObsID~0804550101 for the western shell and from ObsID~0841660301 for the eastern shell containing the central SB. To get our spectral extraction regions, we used the SAS GUI utility \texttt{xmmselect}.

\subsection{Supplementary data}
Our primary analysis of the \spbub{} superbubble region was conducted through X-ray observations, but other data were also available. We therefore used supplementary data in the optical, infrared and radio to understand the multi-wavelength morphology of the object. While the X-ray data traces the hot gas filling the interior of the superbubble, cooler material found within and beyond the supershell is well-distinguished using H$\alpha$, infrared and 21 cm \ion{H}{I} radio images.

\subsubsection{Optical}
Data from the Magellanic Cloud Emission Line Survey (MCELS) observations \citep{mcels1999-smith} were used in our analysis. MCELS was carried out using a 0.6 m University of Michigan/Cerro Tololo Inter-American Observatory (CTIO) Curtis Schmidt Telescope. The entirety of the LMC was mapped using narrow-band filters. In our analysis of the \spbub{} superbubble, we focus on cutouts made from the mosaic primarily using the H$\alpha$ image to trace the supershell. These survey data provided are flux-calibrated with units $10^{-15}$ erg cm$^{-2}$ s$^{-1}$ and continuum subtracted.

We also made use of the more recent Dark Energy Camera Magellanic Clouds Emission Line Surveys (DeMCELS) dataset \citep{Points2024}. The DeMCELS survey comprises multiple programs that used the Dark Energy Camera (DECam) with narrow band filters covering the \ha\ and \sii\ emission-lines, achieving flux limits of $\sim10^{-18}$~erg~s$^{-1}$~cm$^{-2}$. We retrieved the narrow band line and broadband r filter data from the \spbub{} region from the archive\footnote{See \url{https://datalab.noirlab.edu/demcels.php} for details}. We mosaicked the individual panels using the Astropy affiliated \texttt{reproject} package\footnote{https://reproject.readthedocs.io/en/stable/}. We used a subset of the resulting images to estimate the median FWHM of the point sources assuming a 2D Gaussian. We then generated matched PSFs which were used to convolve the two images with the smaller FWHM to the resolution of the image with the largest FWHM. We then performed a continuum subtraction following the steps outlined in Sect.~4 of \citet{Points2024} to produce emission line images of the region. 
        
\subsubsection{Infrared}
We made use of the \emph{Spitzer} \citep{werner2004spitzer} Space Telescope "Surveying the Agents of a Galaxy's Evolution" (SAGE) dataset of the LMC \citep{sage2006}. We gathered \emph{Spitzer}-IRAC \citep[InfraRed Array Camera,][]{fazio2004irac} images at 3.6, 4.5, 5.8 and 8.0 $\mu$m and MIPS \citep[Multiband Imaging Photometer for Spitzer,][]{rieke2004mips} images at 24 $\mu$m. IRAC consists of a four-channel camera which observes two fields of view simultaneously. All four detector arrays are 256 $\times$ 256 pixels in size with a square pixel scale of 1.2". MIPS images at 24 $\mu$m have a 5' square field of view with a detector size of 128 $\times$ 128 pixels. Cutouts (0.5 deg in size) from the larger mosaics centred on the \spbub{} superbubble region were taken from the NASA/IPAC Infrared Science Archive (IRSA)\footnote{https://irsa.ipac.caltech.edu/data/SPITZER/SAGE/overview.html}.

\subsubsection{Radio}\label{sec:radio_data}
We use 21 cm \ion{H}{I} data from the Australian Telescope Compact Array \citep[ATCA\footnote{https://www.narrabri.atnf.csiro.au/},][]{Frater1992ATCA} with data from the \cite{kim_1998_HILMCpeakmap} survey to trace the cold, neutral hydrogen gas surrounding the \spbub{} complex. ATCA is an earth-rotation aperture synthesis radio interferometer which consists of an array of six 22 m antennas capable of imaging in frequencies between 1.1 GHz and 105 GHz with wavelengths from 27 cm down to 3 mm. This data is shown in Fig. \ref{fig:radiosgs7}.

\section{Analysis and results}
\label{sec:sec3}
\subsection{X-ray spectral analysis}
We used the spectral-fitting package XSPEC version 12.13.1 along with its Python implementation PyXspec\footnote{https://heasarc.gsfc.nasa.gov/xanadu/xspec/python/html/} \citep{pyxspec} to perform the spectral extraction and subsequent model-fitting. Abundance tables of \cite{wilms}, photoelectric absorption cross-sections of \cite{balucinska}, and atomic data from ATOMDB 3.0.9\footnote{http://www.atomdb.org/index.php} were used. To allow for the use of the $\chi^2$ statistic in the fitting process, all spectra were rebinned with a minimum of 25 counts per bin.

Analysis of extended, diffuse X-ray regions is often complex and requires a careful treatment of background emission, with instrumental, particle, and astrophysical origins. For the spectral analysis we follow closely the procedure described in \cite{Maggi_LMC}. We extracted a background spectrum and simultaneously fit the science and background spectra over the energy range 0.4 -- 10.0 keV. 

The X-ray background includes the astrophysical X-ray background (AXB) and the QPB. The Local Hot Bubble (LHB), the hot galactic halo (HGH) and unresolved background active galactic nuclei (AGN) all contribute to the AXB. The model components of the AXB are discussed in detail in \citep{snowden2008catalog}. We used two thermal components in collisional ionisation equilibrium (CIE) \citep[APEC,][]{smith2001collisional} to model the unabsorbed LHB and the absorbed HGH along with an absorbed power law for the unresolved AGN. The LHB model temperature was fixed at $kT = 0.1$ keV. The warmer absorbed HGH component was left unconstrained and achieved a best-fit value of $kT=0.196$ keV. We froze the value for the Galactic hydrogen column density $N_H$ at 0.85 $\times~10^{21}$ cm$^{-2}$ obtained using the online HEASARC $N_H$ tool\footnote{https://heasarc.gsfc.nasa.gov/cgi-bin/Tools/w3nh/w3nh.pl} using \ion{H}{I} maps from \cite{dickey1990hi}. Our background AGN component was fixed at $\Gamma$=1.46 and a normalisation of 9.69 $\times~10^{-5}$ photons keV$^{-1}$ cm $^{-2}$ s$^{-1}$ at 1 keV \citep{agnpow_chen} as per the ESAS\footnote{https://www.cosmos.esa.int/web/xmm-newton/xmm-esas} guidelines.

The QPB was modelled together with the contribution from the soft proton contamination (SPC) as described in \cite{Maggi_LMC}. We constrained the QPB using FWC data. The QPB models were fitted to the FWC spectra from the event lists generated in Sect. \ref{sec:particle-bkg}. These best-fit QPB background parameters were then frozen when fitting the science data. We allowed a multiplicative scaling factor and the instrumental line energies/normalisations for the instrumental background to vary in our fits for the combined background effects. Instrumental fluorescence lines within the QPB, such as the Al K line at 1.49 keV and Cu K at 8.05 keV can vary with time and this was accounted for in our models. From our fits we found that the SPC for ObsID~0841660301 was particularly high and dominated the astrophysical background components.

\subsection{Spectral fits}
\begin{figure}[htb]
\begin{adjustwidth}{-2cm}{-2cm}
    \centering
    \begin{subfigure}{0.5\textwidth}
    \includegraphics[clip, trim=1cm 1.25cm 1.5cm 2.5cm, width=\textwidth]{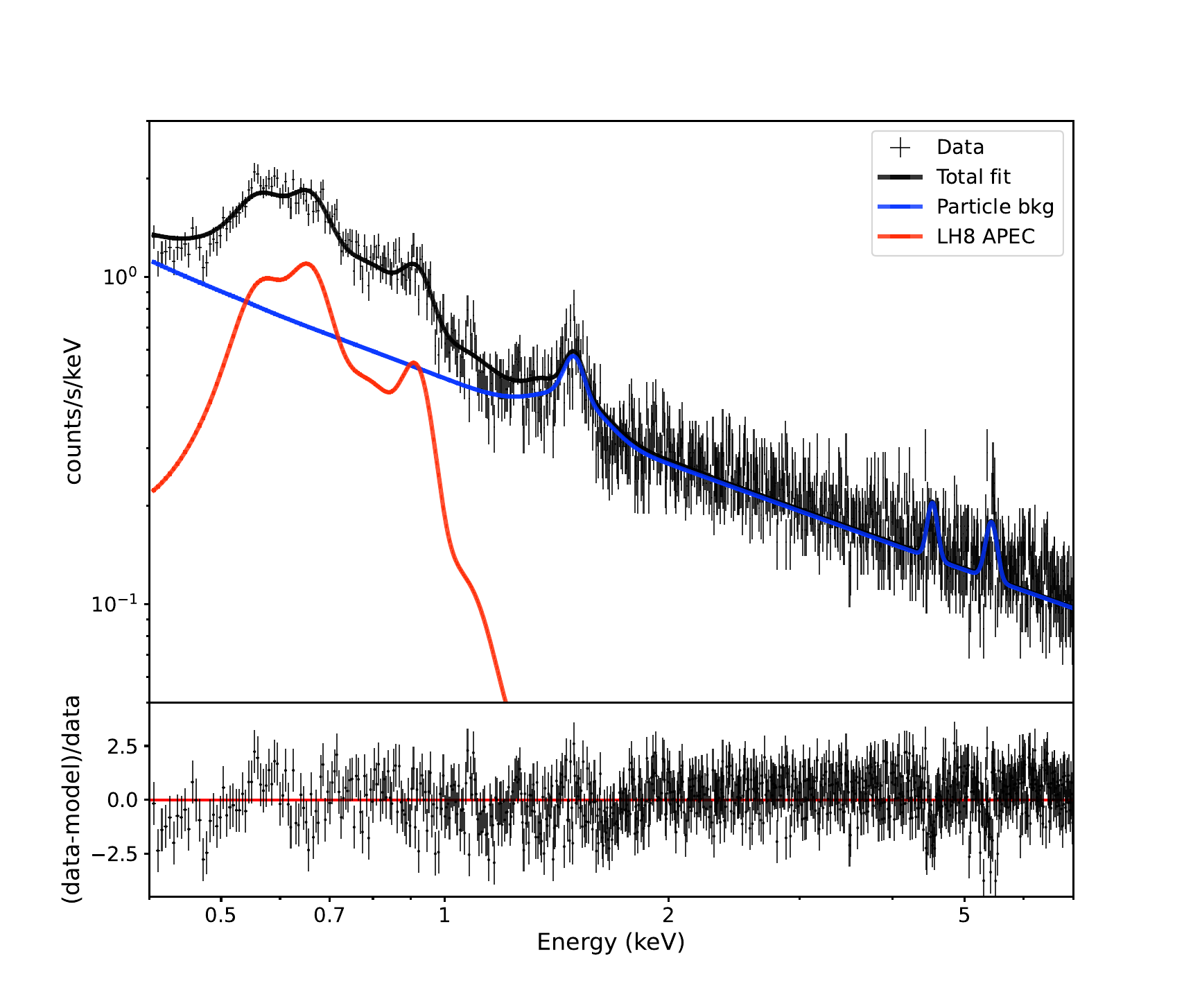}
    \label{fig:model-pn}
    \end{subfigure}
    \begin{subfigure}{0.5\textwidth}
        \includegraphics[clip, trim=1cm 1.25cm 1.5cm 2.5cm, width=\textwidth]{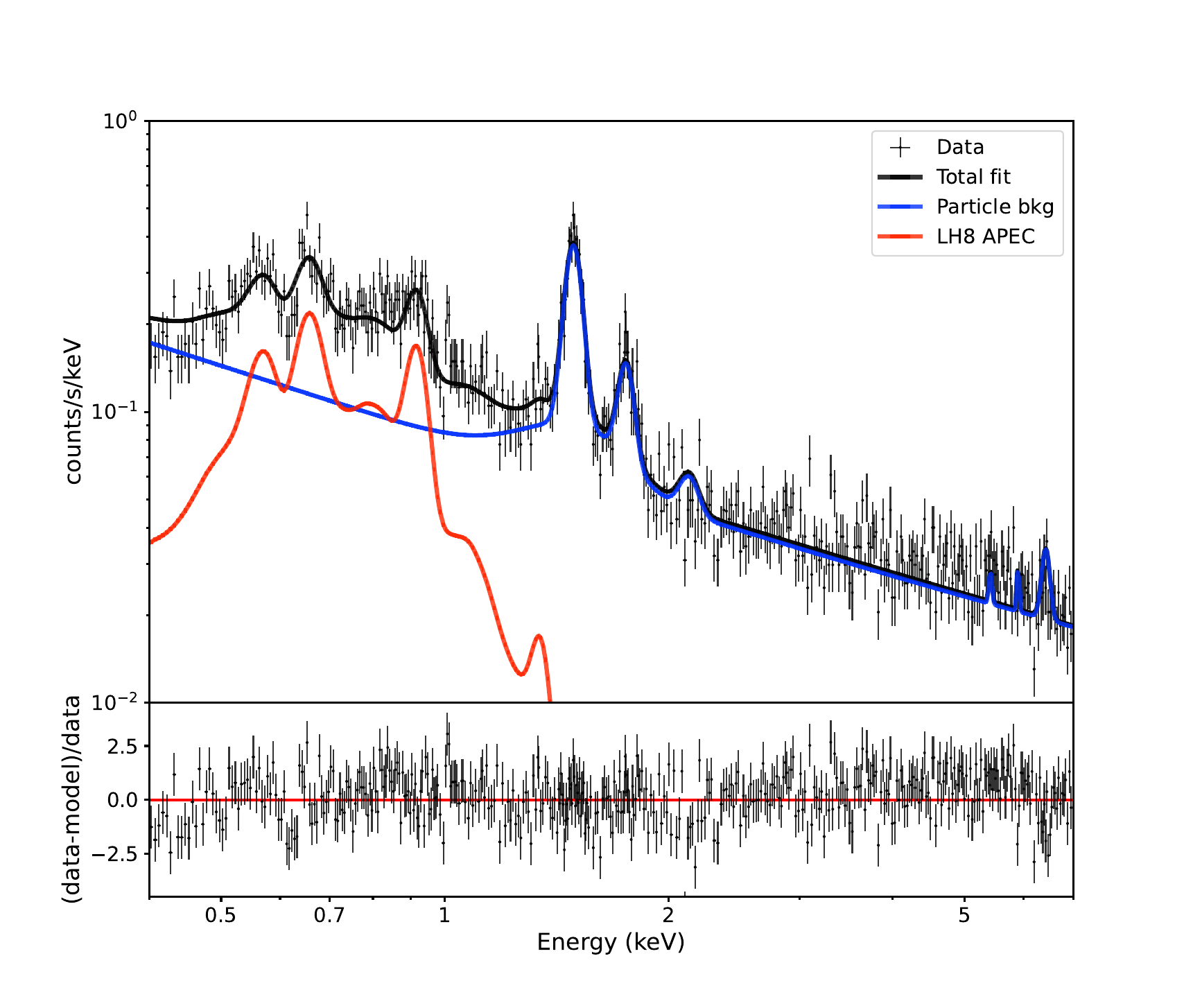}
        \label{fig:model-m2}
    \end{subfigure}
\end{adjustwidth}
    \caption{Best fit APEC model on the EPIC-pn (top) and EPIC-MOS2 (bottom) spectra of the eastern shell region. The total model is shown by the solid black line, the superbubble APEC model is shown in red, and the combined particle background components shown by the blue solid line. Note the LHB and HGH background components were not formally required in the fits because of the high SP contamination in ObsID~0841660301 and were therefore omitted from the fits.}
\label{fig:modelfit}
\end{figure}

We fitted the EPIC-pn and EPIC-MOS2 spectra for each of the eastern and western shell regions simultaneously and with their respective background spectra. Fits to the eastern shell region are shown in Fig.~\ref{fig:modelfit}. Background parameters were linked with a scaling factor to account for the difference in extraction area. We found that it was difficult to constrain the thermal astrophysical background components in spectra extracted from ObsID~0841660301 given the high SPC. Indeed, the lower bound of the confidence intervals for the model normalisations was 0, indicating that the components were not formally required. We therefore omitted them from fits to the SB spectra extracted from ObsID~0841660301.

We fitted each region with an absorbed soft thermal plasma component (APEC, \citealt{smith2001collisional}) in CIE with abundances fixed at the LMC metallicity value of 0.5 $Z/Z_\odot$, giving a reduced $\chi^2$ of 1.28. We performed additional trial fits with the abundance parameter free, with individual abundances such as Fe and O allowed to vary, and also with non-equilibrium ionisation models, but these did not improve the fits and parameters were poorly constrained. We therefore adopted the APEC model at LMC abundances for the \spbub{} eastern and western shell regions. The results of our fits are given in Table \ref{tab:fitresultsmerged}.

A model flux of $F_{X,(0.3-10.0)}=$ 1.15 $\times~10^{-11}$ ergs cm$^{-2}$ s$^{-1}$ was calculated for the eastern region. We determine the emission measure (EM) of the eastern shell region using Eq. \ref{eq:em} for the normalisation;

\begin{equation} \label{eq:em}
\begin{split}
    norm_{SB} & = \frac{10^{-14}}{4\pi D^2}\int_V n_e n_H dV \\
    EM_{SB} & = \int_V n_e n_H dV \\
    \rightarrow EM_{SB} & = norm_{SB} * \frac{4\pi D^2}{10^{-14}}
\end{split}
\end{equation}
with $n_e$ and $n_H$ being the electron and hydrogen densities of the emitting plasma respectively. Using the known distance to the LMC of 50 kpc \citep{pietrzynski2019distance}, the EM is estimated to be $63.4(45.2-81.7)\times 10^{58}$ cm$^{-3}$.

\begin{table}[htb]
    \centering
    \caption{Spectral parameters gathered from the fits of the eastern and western shell regions.}
    \begin{tabular}{lll}
    \toprule
    Parameter & Eastern shell & Western shell \\
    \midrule \midrule
    $N_{H}$ (10$^{22}$ cm$^{-2}$) & 0.56 (0.50--0.61) & 0.61 (0.55--0.69) \\
    $kT$ (keV) & 0.17 (0.16--0.18) & 0.17 (0.15--0.17)  \\
    $norm$ (10$^{-4}$) & 212 (151--273) & 124 (93.3--202) \\
    $F$ ($10^{-11}$ ergs cm$^{-2}$ s$^{-1}$) & 1.15 & 1.29 \\
    $EM$ ($10^{58}$ cm$^{-3}$) & 63.4 (45.2--81.7) & 37.1 (27.9--60.5) \\ 
    \midrule
    \multicolumn{3}{l}{Background parameters} \\
    \midrule
    $norm_{~\mathrm{LHB}}$ (10$^{-4}$) & --  & 4.83 (4.08--5.39) \\
    $kT_{~\mathrm{HGH}}$ (keV) & --  & 0.30 (0.27--0.32) \\
    $norm_{~\mathrm{HGH}}$ (10$^{-4}$) & -- & 1.19 (1.00--1.44) \\
    \midrule
    $\chi^{2}_{\nu}$ (d.o.f.) & 1.28 (1582) & 1.27 (804) \\
    \bottomrule
    \end{tabular}
    \tablefoot{Source and background spectra were fit simultaneously. The quoted X-ray flux $F$ is over the 0.3 - 10.0 keV range and is de-absorbed. Because of high SPC in ObsID~0841660301, the thermal background components were not formally required for the SB region and were therefore omitted from the fits. See text for details.}
    \label{tab:fitresultsmerged}
\end{table}

\begin{table}[htb]
    \centering
    \caption{Derived physical parameters for the superbubble regions discussed in Sect. \ref{sec:specfitresults}.}
    \begin{tabular}{lcc}
    \toprule
    Parameter & Eastern shell & Western shell \\
    \midrule
    $V$ ($10^{61}$ cm$^3$)& $1.96~(\pm0.01)$ & $2.90~(\pm0.01)$  \\
    $n_H$ (cm$^{-3}$) & $0.16~(\pm 0.04)$ & $0.10~(\pm 0.05)$ \\
    $P/k$ ($10^{5}$ K cm$^{-3}$) & $7.45~(\pm 1.96)$ & $4.69~(\pm 2.07)$ \\
    $E$ ($10^{51}$ erg) & $3.02~(\pm 0.79)$ & $2.81~(\pm 1.24)$ \\
    $M$ ($10^{3}$ M$_\odot$) & $3.79~(\pm 0.97)$ & $3.53~(\pm 1.55)$ \\
    \bottomrule
    \end{tabular}
    \label{tab:bubble_params}
\end{table}

\subsection{Properties of the X-ray gas}
\begin{figure*}
\sidecaption
    \includegraphics[width=12cm]{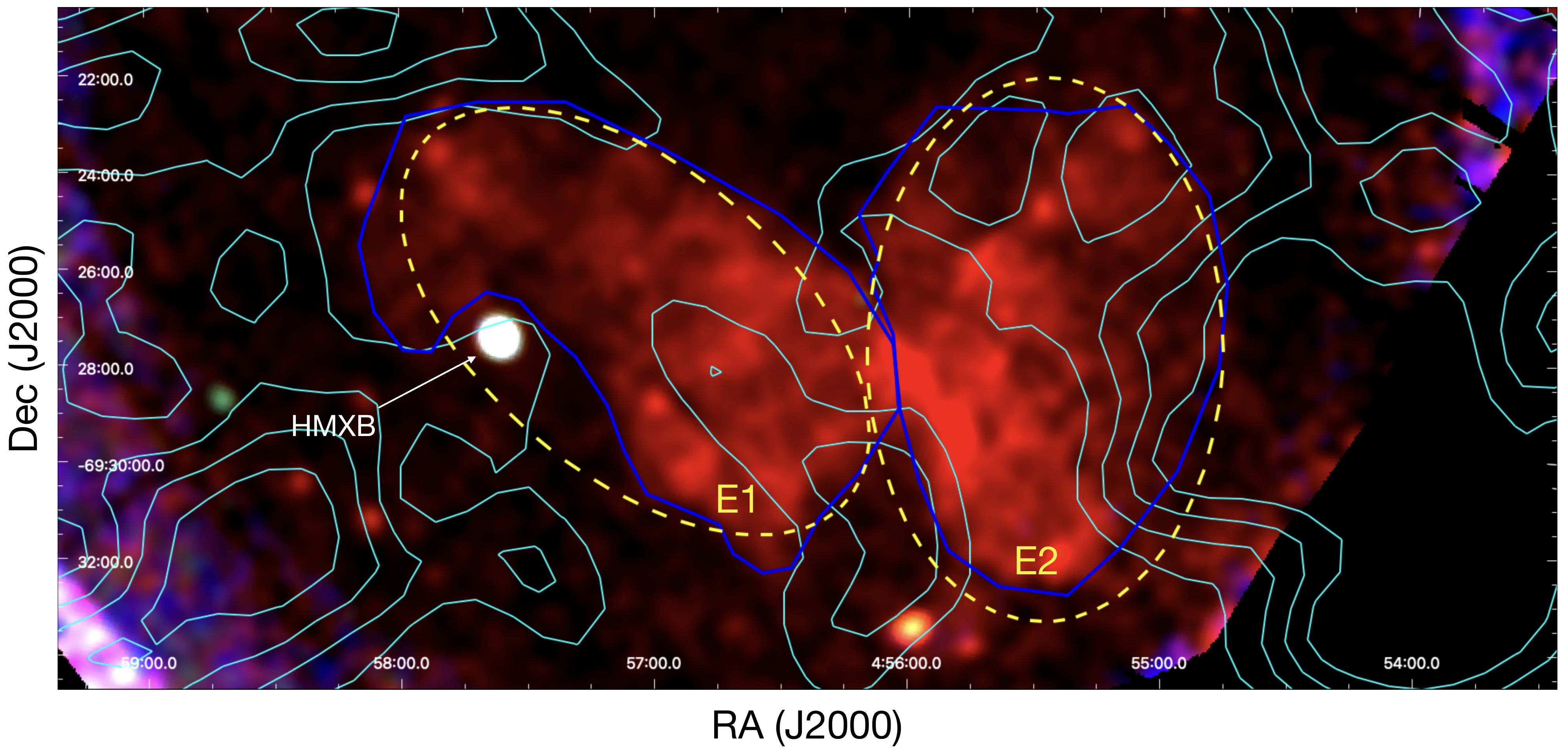}
    \caption{Three colour X-ray image of the \spbub{} SB with red, green, and blue corresponding to the 0.2 - 0.5 keV, 2.0 - 4.5 keV, and 4.5 - 12.0 keV bands respectively. Overlaid in cyan are \ion{H}{I} 21~cm contours from the ATCA. In yellow, the E1 and E2 ellipsoids used for volume estimation. The extraction regions used for spectra generation are also overlaid in blue. The HMXB studied in \cite{haberl2022binaries} is also labelled.}
    \label{fig:xray-3colour}
\end{figure*}

\label{sec:specfitresults}
We estimated the physical properties of this X-ray emitting gas following the methods of \cite{sasaki2011xmm} and \cite{kavanagh2012xmm}. Due to its complex morphology, we determine the total volume of the MSFR using two ellipsoids: an eastern (E1) and western ellipsoid (E2) (see Fig. \ref{fig:xray-3colour}) fitted to our image using the \texttt{photutils} package \citep{larry_bradley_2024_10967176}. E1: $a=(338''\pm 15'')=(81.9\pm 3.6)$ pc, $b=(182''\pm 15'')=(44.1\pm 3.6)$ pc and E2: $a=(338'' \pm 15'')=(81.9\pm3.6)$ pc, $b=(221''\pm 15'')=(53.6\pm 3.6)$ pc. Using $V=(4/3)\pi abc$ and the assumption that the SB is prolate and is symmetrical around one axis ($a>b=c$), we combine the ellipsoid volumes $V_{\text{prolate}}=V_1+V_2 = (4.86\pm 0.01)\times10^{61}$ cm${^3}$. We can also assume the superbubble volume is oblate ($a=c>b$), which gives us a total volume $V_{\text{oblate}}=(8.06 \pm 0.01)\times 10^{61}$ cm$^3$. We derive the subsequent physical parameters for the eastern/western regions with $V_{\text{east}}=(1.96\pm 0.01)\times 10^{61}$ cm$^3$ and $V_{\text{west}}=(2.90\pm 0.01)\times 10^{61}$ cm$^3$ respectively, assuming the volumes to be prolate. The hydrogen density is calculated using the normalisation of the \texttt{apec} component of our model along with the metallicity of the LMC $\zeta_{LMC}=0.5$, $n_e=(1.2+0.013\zeta_{LMC})$ which gives $n_e\approx 1.21n_H$. The normalisation of the spectral fit is modified with a gas filling parameter $f$ below in Eq. \ref{eq:norm}; 
\begin{equation} \label{eq:norm}
\begin{split}
    norm_{SB} & = \frac{10^{-14}}{4\pi D^2}\int_V n_e n_H f dV \\
    & \approx \frac{10^{-14}}{4\pi D^2} 1.21 n_H^2 f V \\ 
    & = 4.04 \times 10^{-62} n_H^2 f V \quad [\text{cm$^{-5}$}].
\end{split}
\end{equation}
where $V$ represents either $V_{\text{east}}$ or $V_{\text{west}}$ with the normalisations $norm_{SB}$ for those respective regions. We assume a filling parameter of 1 with the bubble being almost fully filled with hot X-ray emitting gas. Rearranging for the hydrogen gas density $n_H$ gives:
\begin{equation}\label{eq:density}
\begin{split}
    n_H & = 4.97 \times 10^{30} \sqrt{\frac{norm_{SB}}{fV}} \quad [\text{cm}^{-3}] \\
    & = (0.16 \pm 0.04) \quad \text{cm}^{-3},
\end{split}
\end{equation}
for the eastern shell and, $n_H=0.1 \pm 0.05$ cm$^{-3}$ for the western shell. 

We estimate the pressure of the gas within the superbubble at a temperature of $kT=(0.17 \pm 0.01)$ keV using the density $n_H$ calculated above. We assume the pressure is constant throughout the entire volume:
\begin{equation}\label{eq:pressure}
    \begin{split}
        P/k & = (n_e + 1.1n_H)T = 2.31n_HT \\
        & = (7.45 \pm 1.96) \times 10^5 \quad \text{K}~\text{cm}^{-3},
    \end{split}
\end{equation}
for the eastern region and $(4.69 \pm 2.07) \times 10^5$ K cm$^{-3}$ for the western region. \cite{kim2023tigress} used TIGRESS-NCR simulations to model the star-forming ISM across various galactic conditions. They estimate a total mid-plane pressure, which includes the hot ISM found in star-forming regions such as SBs, to be $P/k=10^3 - 10^6$ cm$^{-3}$ K. Therefore our pressure value is in agreement with these simulations. Our pressure values are similar to that of the N 206 superbubble described in \cite{kavanagh2012xmm} to be $\approx 5\times 10^5$ cm$^{-3}$ K and about half of the pressure in N 158 described in \cite{sasaki2011xmm}.

Thermal energy content within the superbubble is given by $E_{\text{SB,east}}=3/2PV_{\text{east}}=(3.02\pm 0.79) \times 10^{51}$ erg for the eastern region and $E_{\text{SB,west}}=3/2PV_{\text{west}}=(2.81\pm 1.24) \times 10^{51}$ erg for the western shell. The mass inside a homogeneous bubble is also given by $M_{\text{SB,east/west}}=2.31n_H\times\mu\times m_H \times V_{\text{SB,east/west}}$, where $\mu=0.61$ is the mean molecular weight of a fully ionised gas and $m_H$ is the hydrogen mass. We find the mass of the X-ray emitting gas within the \spbub{} SB to be $M_{\text{SB,east}}=(3800\pm 970)$ M$_{\odot}$ and $M_{\text{SB,west}}=(3530\pm 1550)$ M$_{\odot}$ for the eastern and western shells respectively. These values are listed in Table \ref{tab:bubble_params}.

\section{Discussion}
\label{sec:sec4}
\subsection{\spbub{} morphology}
\label{sec:sb-morphology}
The morphological comparison between X-ray, optical and infrared wavelengths is shown in Fig.~\ref{fig:multi-wavelength} with red, green and blue corresponding to 24 $\mu$m MIPS, DeMCELS H$\alpha$, and 0.3 - 1.0 keV EPIC images, respectively. The eastern X-ray shell region is mostly bounded by a bright H$\alpha$. We consider this to be the \spbub{} SB. There are two notable blowouts on the eastern and western sides, evident by the broken H$\alpha$ shell with X-ray emitting gas extending into these regions. The western blowout breaks into the western shell region which is quite bright in X-rays. The southern edge of the SB exhibits a possible emerging blowout, labelled the southern blowout in Fig.~\ref{fig:multi-wavelength}, with the partial fragmentation of the H$\alpha$ shell structure, though much less significant than the others. Unlike many other LMC SBs, there is no clear powering cluster for this SB. There are two smaller embedded stellar clusters along the northern edge of the regions, NGC~1767 and NGC~1782, both indicated on Fig.~\ref{fig:multi-wavelength}. They are discussed in more detail in Sect. \ref{sect:mech_input_1767}, including their likely role as powering sources at the current epoch.

We compared the X-ray morphology to the 21 cm ATCA image of the \spbub{} region tracing \ion{H}{I} in Fig. \ref{fig:radiosgs7}. This clearly shows how the stellar feedback from the MSFR has shaped the cold medium. The western shell appears to be bounded by high \ion{H}{I} emission along its western edge. It is unclear where this \ion{H}{I} structure lies relative to the X-ray gas in the western shell region. If in the foreground, the western shell may actually extend further but is being absorbed. There are large cavities in the \ion{H}{I} distribution to the north and south of the \spbub{}. These are likely part of the larger SGS7 structure and have been carved out by previous massive stellar populations in the region. 

\subsection{Comparison to other SBs}
\label{sec:sb-comparison}
\begin{figure}
    \centering
    \includegraphics[width=0.5\textwidth]{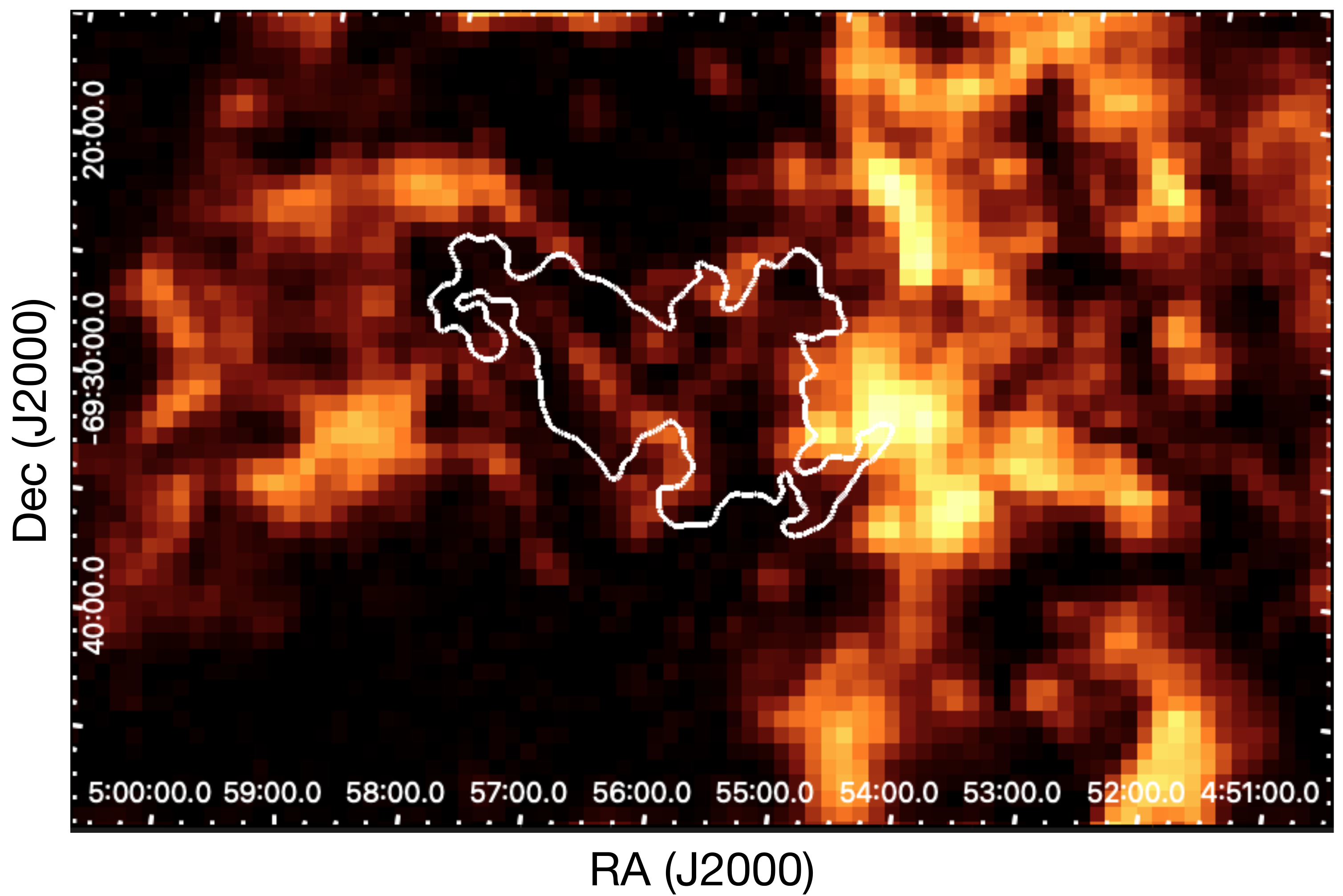}
    \caption{ATCA 21 cm image of the wider LMC-7 supergiant shell with X-ray contours overlaid. An absorption column of \ion{H}{I} to the west could be obstructing further X-ray emission.}
    \label{fig:radiosgs7}
\end{figure}

The \spbub{} SB exhibits a complex morphology and with blowouts and low density channels similar to the well-known Carina \citep{townsley2011chandra} and 30 Doradus (\citealt{chu1990model}; \citealt{wang1991x}) MSFRs. 
We compared the X-ray emission from \spbub{} to these and other Galactic and LMC MSFRs and SBs in the context of their area, luminosities, and surface brightness. The comparative properties are shown in Table \ref{tab:sb-comparison}. The area of the \spbub{} SB was determined to be approximately 23,000 pc$^2$. This is much larger than that of Carina but smaller than the full 30 Doradus region with area values of 1207 and 33,389 pc$^2$ respectively. It is also worth noting the high absorption column density $N_H$ found for \spbub{} is comparable to some SNRs in the LMC such as J0537-6910 and J0540-6920 \citep{maggi2015population} and to the 30 Doradus star-forming complex. This high $N_H$ value could be due to a non-uniform foreground absorbing material around the \spbub{} region (see Fig.~\ref{fig:radiosgs7}). We estimated the surface brightness of the \spbub{} SB, the intrinsic luminosity per unit area, to be $8.35 \times 10^{31}$ erg s$^{-1}$ pc$^{-2}$ which is lower than 30 Doradus but higher than Carina which have surface brightness values of $13.2\times 10^{31}$ erg s$^{-1}$ pc$^{-2}$ and $2.63\times 10^{31}$ erg s$^{-1}$ pc$^{-2}$ respectively. For the smaller 30 Dor C SB, the surface brightness is comparable to the \spbub{} SB with a value of $1.35\times 10^{31}$ erg s$^{-1}$ pc$^{-2}$. Comparing the scale of the \spbub{} region to the smaller 30 Dor C SB suggests it is an older, more evolved SB. Other LMC superbubbles such as N 206 and N 158 both exhibit lower gas density $n_H$ values of 1.9 and $4.0 \times 10^{-2}$ cm$^{-3}$ compared to our values of 0.16 cm$^{-3}$ and 0.10 cm$^{-3}$ in the eastern and western shells, respectively. The density value we find for the \spbub{} SB is more comparable to Galactic objects such as M17 \citep{dunne2008m17} which has a gas density value of 0.2-0.4 cm$^{-3}$ near the centre and increasing outwards with an ambient density of $\sim$60 cm$^{-3}$. \cite{suarez2024galacticsbs3d} performs 3D simulations for two superbubble formation models based on data from our Galaxy with central halo gas density values of 0.1 cm$^{-3}$, agreeing with our value gathered from spectral fitting. The NGC 3079 nuclear superbubble in \cite{taoli2024ngc3079} gives gas density ($n\approx n_e/1.21$) values ranging from 0.12 to 0.27 cm$^{-3}$ depending on the area in the SB. The X-ray luminosity $L_X$ for the \spbub{} SB with a value of $1.93\times 10^{36}$ erg s$^{-1}$ over the 0.5-10.0 keV band is higher than Galactic objects and similar to luminosities of the LMC sample. 

\begin{table*}[htb]
    \centering
    \caption{Comparison between the main properties of different superbubbles.}
    \begin{tabular}{lcccccc}
    \toprule
    Object Name & $D$ & Area & $N_H$ & $kT$ & $L_{X}$ & Surface Brightness\\
                & (kpc) & (pc$^2$) & ($\times 10^{22}$ cm$^{-2}$) & (keV) & ($\times 10^{35}$ erg s$^{-1}$) & ($\log{}$ erg s$^{-1}$ pc$^{-2}$) \\
                &  & \tablefootmark{a} & \tablefootmark{b} & \tablefootmark{c} & \tablefootmark{d} & \tablefootmark{e} \\
    \midrule \midrule
    \spbub{} SB & 50 & $\sim$23000 & 0.59 & 0.17 (CIE) & 19.3 (0.5-7.0) & 31.92 \\
    Carina & 2.3 & 1207 & 0.40 & 0.31 (NEI) & 0.32 (0.5-7.0) & 31.42 \\
    M17 & 2.1 & 85.6 & 0.16 & 0.28 (NEI) & 0.20 (0.5-7.0) & 32.4 \\
    30 Doradus & 50 & 33389 & 0.69 & 0.20 (NEI) & 44.0 (0.5-7.0) & 32.12 \\
    DEM L192\tablefootmark{1} & 50 & 10568 & 0.034 & 0.21 & $\sim$4.0 (0.3-3.0) & 31.57 \\
    N 70\tablefootmark{2} & 50 & 7853 (approx.) & 0.05 & 0.22 & 2.4 (0.2-1.1) & 31.48 \\
    N 185\tablefootmark{2} & 50 & 7853 (approx.) & 0.05 & 0.20 & 2.1 (0.2-1.1) & 31.43 \\
    30 Dor C\tablefootmark{3} & 50 & 7853 & 0.47 & 0.22,0.74 (CIE) & 1.06 (0.5-9.0) & 31.13 \\
    \bottomrule
    \end{tabular}
    \tablefoot{ \\
    \tablefoottext{a}{Geometric area of the extraction region used for spectral analysis.} \\
    \tablefoottext{b}{Density of absorbing hydrogen column.} \\
    \tablefoottext{c}{Plasma temperature from best fitting model.} \\
    \tablefoottext{d}{Total absorption-corrected, intrinsic X-ray luminosity. The numbers in brackets denote the band over which this was calculated in keV.} \\
    \tablefoottext{e}{Surface brightness (intrinsic luminosity per unit area) of the bubble region. The surface brightness values were estimated by averaging over the values in \cite{townsley2011integrated} for the different NEI and Gaussian components for 30 Doradus and Carina. Other surface brightness were estimated from the assumed geometry and X-ray luminosities.} \\
    \tablefoottext{1}{Assuming a spherical geometry. Area taken from radius from \cite{oey1998dem192}. All other parameters taken from \cite{cooper2004energy}}. \\
    \tablefoottext{2}{\cite{reyesN70N185} assume a radius of 50 pc for these superbubbles.} \\
    \tablefoottext{3}{Based on 100 pc diameter from \cite{chi2024thermal}. Absorption and temperatures from \cite{sasaki2022first}. The temperatures quoted are for a two-temperature model ($kT_1$ and $kT_2$ components). Luminosities derived from the average over 3 shells in \cite{bamba2004thermal}.}
    }
    \label{tab:sb-comparison}
\end{table*}

\subsection{Energy budget}

Many studies in the literature have shown that an energy deficit is observed for almost all SBs, i.e. the observed energy content is much less than predicted from stellar and supernova feedback \citep[e.g.,][]{cooper2004energy,maddox2009nonthermal,jaskot2011observational}. In the following we assess if the same is true of the \spbub{} SB and discuss reasons for this.

\subsubsection{Energy stored in optical shell}
We estimated the energy stored in the H$\alpha$ shell using the continuum subtracted MCELS image. The brightest regions of the H$\alpha$ shell which dominate its energy are roughly circular. We therefore defined a circular aperture of radius of 3.6' encompassing these regions and determined the integrated flux, masking any residual stellar contribution. The H$\alpha$ luminosity was then estimated from the flux using the known distance to the LMC to be $L_{\text{H}_\alpha} = 7.62 \times 10^{37}$ erg s$^{-1}$. For ionised gas at temperatures of $\sim10^4$ K, $L_{\text{H}_\alpha} = 3.56\times 10^{-25} n_e n_p V f_{sh}$, where $n_e$ and $n_p$ are the number densities of electrons and protons, $V$ is the SB volume, and $f_{sh}$ is the filling parameter. For singly ionised helium, $n_e = 1.1n_p$ \citep{dunne2001x}. Our aperture radius of 3.6' corresponds to 52.3 pc. We assume a spherical volume giving $1.76 \times 10^{61}$ cm$^3$. Using these values we obtain $n_p = 3.65 f^{1/2}_{sh}$ cm$^{-3}$. The total mass of the shell is given by $M_{sh} = 1.27n_e V f_{sh} m_H = 6.87 \times 10^4$ M$_\odot \, f^{1/2}_{sh}$. Due to the complex nature of the morphology of the region, the fraction that the H$\alpha$ shell occupies is difficult to determine. We conservatively estimate this to be $\sim$ 10 - 15\% of the radius which gives $f_{sh}=0.30 \pm 0.05$. Using this value for $f_{sh}$ we determine the total mass to be $M_{sh} = 3.77(3.44-4.07) \times 10^4$ M$_\odot$\footnote{The numbers in parenthesis are the 90\% confidence interval bounds.}. From the \cite{dunne2001x} sample of superbubbles, the H$\alpha$ expansion velocities for SBs comparable to this one based on $L_{\text{H}_\alpha}$ are in the range 20 -- 45 km s$^{-1}$. We use these velocity estimates to calculate the energy stored in the optical shell using $E_{sh}=\frac{1}{2} M v_{\text{exp}}^2$, deriving an upper limit of $E_{sh} = 7.58~(6.92-8.19) \times 10^{50}$ erg and a lower limit of $1.50~(1.37-1.62) \times 10^{50}$ erg. 

\subsubsection{Energy stored in \ion{H}{I} shell}
As the supershell expands it also causes surrounding \ion{H}{I} to be swept outwards. The mass of this \ion{H}{I} gas can be estimated using 21~cm emission line data (Sect.~\ref{sec:radio_data}). We used the \ion{H}{I} peak temperature image to determine the flux density towards the \spbub{} superbubble. Since the \ion{H}{I} shell is expected to be larger than the H$\alpha$ shell, we used an aperture 20\% larger than the supershell and measured a flux density of 9.81~Jy, which corresponds to a mass of $\sim9.5\times10^{3}$~M$_{\sun}$ at the distance of the LMC. Assuming the same expansion velocity as the H$\alpha$ shell, this corresponds to a kinetic energy of $\sim1\times10^{50}$~erg. We have made several assumptions in this calculation. We used the \ion{H}{I} peak temperature map, some of the \ion{H}{I} emission is likely not associated with the superbubble, and we assumed a fixed expansion velocity at the upper end of the H$\alpha$ shell estimate. Therefore, this value should be considered an upper limit.

\begin{figure*}[!t]
    \centering
    \includegraphics[width=0.9\textwidth]{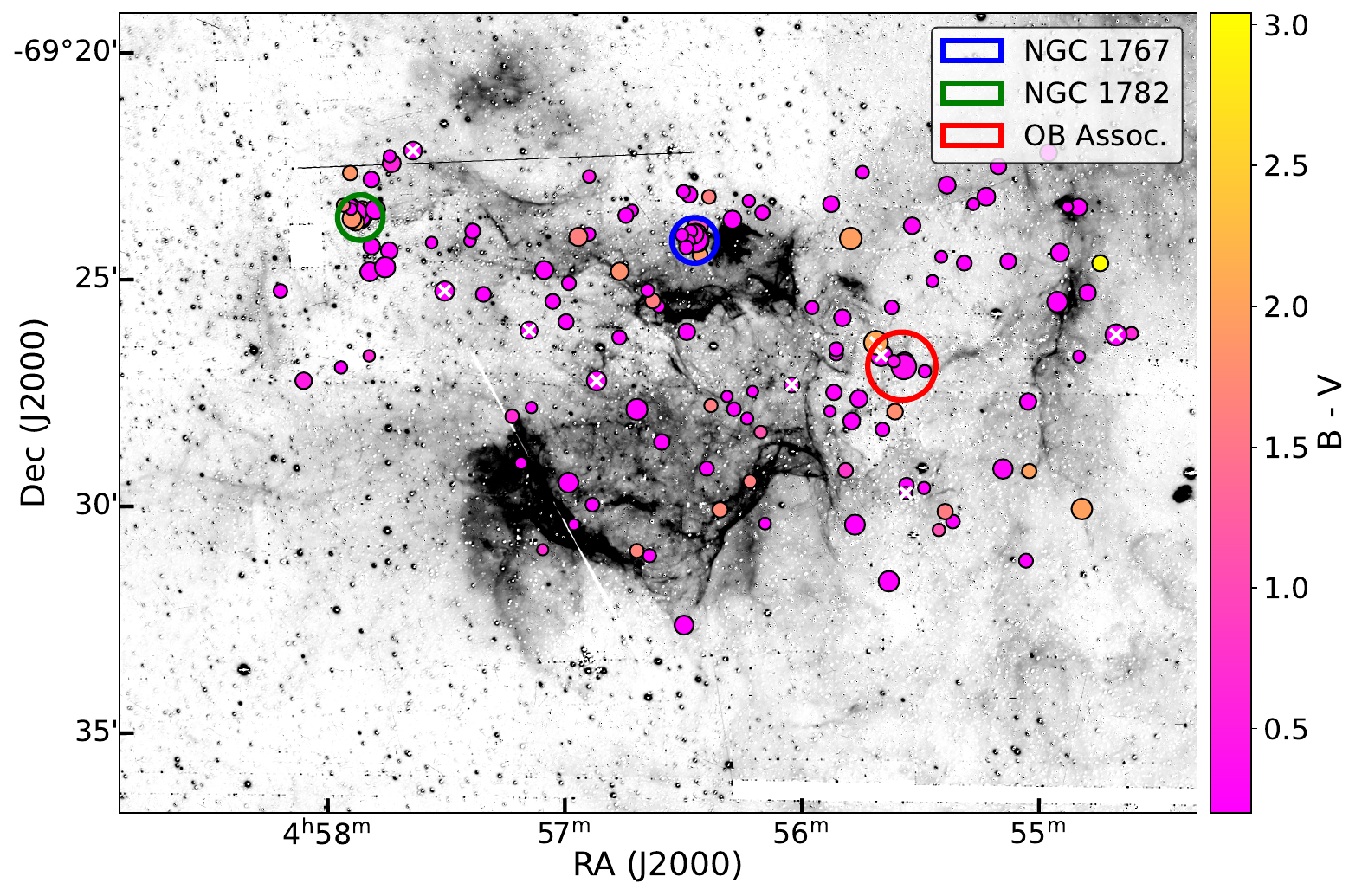}
    \caption{Bright stars from the photometry catalogue of \cite{zaritsky2004magellanic} plotted on the continuum-subtracted H$\alpha$ DeMCELS image. The size of the plotted sources indicates the $V$ band magnitude, with sizes linearly scaled between the smallest at $V=15.0$ and largest at $V=11.2$. The colour of the data points indicates the $B-V$ colour with scaling shown shown by the colour bar. Identified post-main sequence stars are indicated by white `$\times$' symbols. The location of two known clusters NGC~1767 and NGC~1782 shown by blue and green circles, respectively, both with diameter of 1~arcmin, and the OB association shown by the red circle of diameter 1.5~arcmin. }
    \label{fig:bright_stars}
\end{figure*}

\subsubsection{Stellar content inside \spbub{} SB}
Unlike other LMC SBs, there is no obvious massive stellar cluster that can be identified as the powering source. We therefore used the catalogue of stellar photometry data from \cite{zaritsky2004magellanic} to assess the stellar population in the region. We limited our search to sources in and near the X-ray bright regions. We consider only sources with $V<15$, which corresponds to $\gtrsim10$~M$_{\sun}$ stars on the main sequence at the distance of the LMC. We removed all sources with GAIA DR3 counterparts with distances inconsistent with the LMC. This resulted in 144 sources within our search region. These sources are plotted in Fig.~\ref{fig:bright_stars}. 
It is important to note that it is difficult to determine whether these remaining sources do indeed belong to the LMC and \spbub{} regions, or are foreground or background stars without a GAIA counterpart.

We tried to identify foreground stars but this was difficult due to the absence of high-resolution optical data. We searched for source identifications in the SIMBAD database. Several of the sources were identified as high-mass stars in the post-main sequence. These are also indicated in Fig.~\ref{fig:bright_stars}. Many of the sources are dispersed throughout the \spbub{} region, though three groupings are clear, two along the northern edge of the H$\alpha$ shell, identified as the NGC~1767 (RA (J200) = 4h 56m 27.46s, Dec = -69$\degree$ 24' 10.2") and NGC~1782 (RA (J2000) = 4h 57m 51.21s, Dec = -69$\degree$ 23' 39.6") clusters, and an OB association (RA = 4h 55m 35.32s, Dec = -69$\degree$ 26' 55.6") in the western bubble region, which also contains massive post-main sequence stars (see Fig.~\ref{fig:bright_stars}). 

\begin{figure*}
\sidecaption
    \includegraphics[width=12cm]{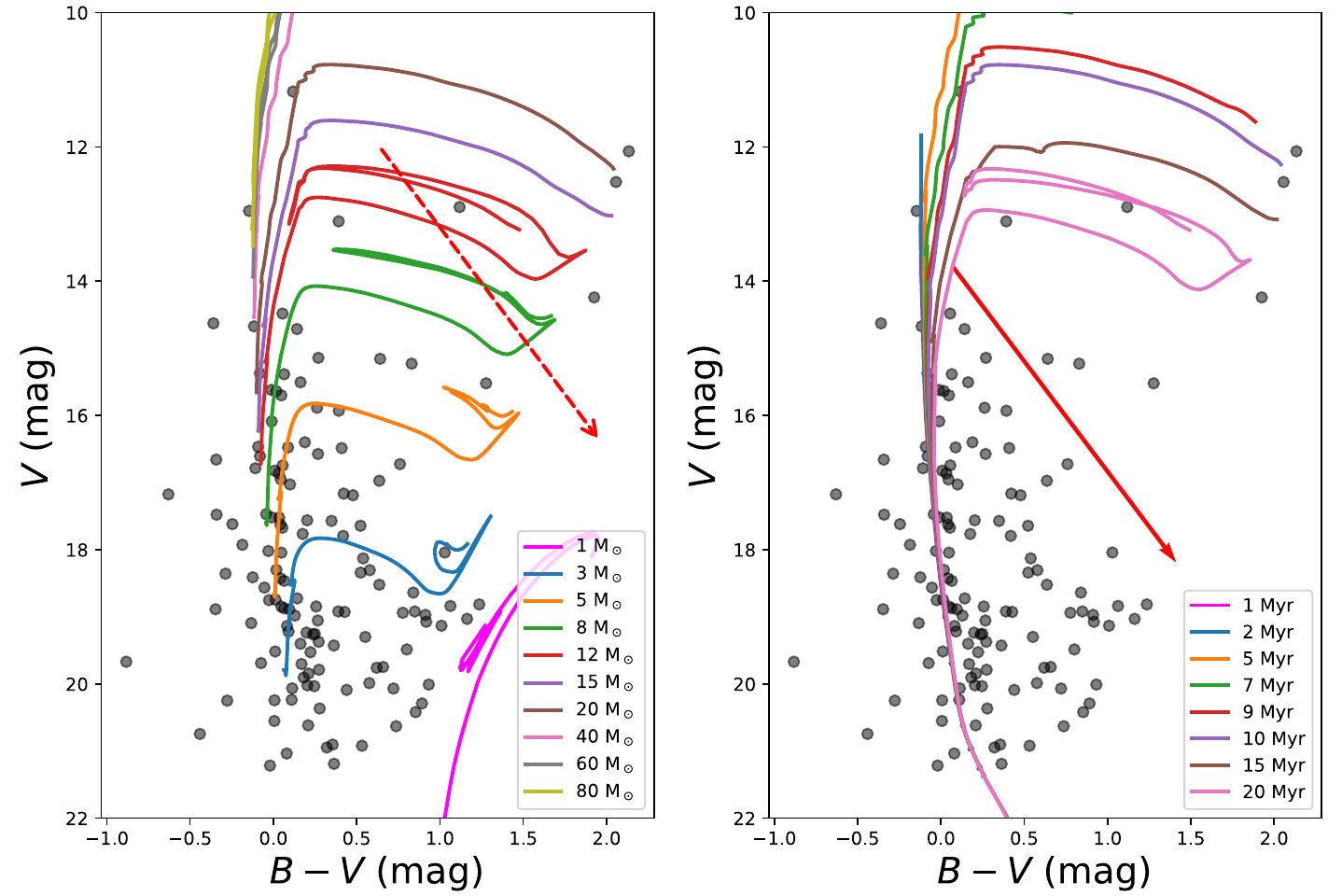}
    \caption{CMD with apparent \textit{V} versus (\textit{B-V}) for the NGC~1767 cluster located at the northern shell of the \spbub{} superbubble. The photometric data are from \cite{zaritsky2004magellanic}. \textbf{Left}: Evolutionary tracks (MS to CHeB phases only) for varying masses ranging from 1~M$_\odot$ to 80~M$_\odot$. \textbf{Right}: Fitted isochrones from the MIST suite are overlaid with varying stellar ages from 1-20~Myr. The reddening vector corresponding to a $A_V=5.0$ shift is shown as a red arrow for both plots.}
    \label{fig:CMDs}
\end{figure*}

We investigated the stellar content of these regions to determine their ages, initial masses, and also to estimate their mechanical output to compare to the observed energies estimated above. Colour magnitude diagrams (CMDs) were produced from the \cite{zaritsky2004magellanic} photometric catalogue for each with search radii of 0.5~arcmin for the clusters and 1.0~arcmin for the OB association (see Fig.~\ref{fig:bright_stars}). Isochrones and stellar evolutionary tracks were then fitted from the MESA Isochrones \& Stellar Tracks (MIST)\footnote{https://waps.cfa.harvard.edu/MIST/index.html} suite (\citealt{mist0}; \citealt{mist1}). The photometric data were fitted using an Fe/H ratio of -0.37 dex based on the average LMC photometric metallicity from \cite{choudhury2016photometric}, with dust extinction $A_V$ values ranging from 0.1 - 1.0 from studies by \cite{furuta2022extinction} and \cite{imara2007extinction}. This allowed us to estimate the age and masses of the stellar populations in each region. Examples are shown for the NGC~1767 in Fig. \ref{fig:CMDs}.

\subsubsection{NGC~1767}
\label{sect:mech_input_1767}
Previous photometric studies of NGC~1767 have been conducted by \cite{kumar2008ccd}. They estimated the age of the cluster to be $23\pm 3$ Myr. To be consistent with our other clusters, we performed our own photometric study of NGC~1767. As stated above, despite cleaning the data for foreground sources, we cannot be sure that some remain. We continue under the assumption that the stars are in \spbub{} which at least allows an upper limit to be determined for the mechanical input from the stellar population.

We estimate the age of the NGC~1767 cluster to be 10-20~Myr using the MIST isochrones in the CMD, consistent with \cite{kumar2008ccd}. In order to determine the initial cluster mass, we assume that the stellar population was formed in an instantaneous burst of star formation and follows the log-normal Chabrier initial mass function \citep[IMF,][]{chabrier2003}. We calibrated the IMF to our stellar population by determining the number of stars in the 3-5~M$_\odot$ range, with uncertainty being the Poisson count error, and scaling the IMF appropriately. We then integrated the IMF giving an initial cluster mass of $1390~(\pm188$)~M$_\odot$. We note that, as outlined by \cite{kavanagh2012xmm}, the non-application of binary evolutionary tracks in our analysis adds additional uncertainty. 

To determine the stellar mechanical output of NGC~1767, we used the \emph{Starburst99}\footnote{https://www.stsci.edu/science/starburst99/docs/default.htm} stellar population synthesis code \citep{leitherer1999starburst, leitherer2014starburst}, which predicts the total energy, mass loss rates, and the supernovae (SNe) rates. We used stellar evolution models of the Geneva group with rotation $v/v_{\text{crit}}=0.4$ and metallicity 0.014 $Z/Z_\odot$. We assume an initial burst of star formation of a cluster with a mass of 1390 M$_\odot$, as determined above. \emph{Starburst99} predicts that the cluster has injected $(6.07 - 10.0) \times 10^{51}$ erg and 117--167~M$_\odot$ over its estimated lifetime of $\sim$10-20 Myr. This comprises contributions from both stars and SNe. We integrated the SNe rate, finding that $\sim3-7$ SNe have occurred in the cluster.

\subsubsection{NGC~1782 and OB association}
We followed the same approach as described in Sect.~\ref{sect:mech_input_1767} to determine the mechanical input to the \spbub{} region from NGC~1782 and the OB association in the western bubble. We estimate an initial cluster mass of 2316 M$_\odot$ with injection values of $(13.68 - 16.71)\times 10^{51}$ erg and 243-279 M$_\odot$ over an estimated lifetime of $\sim$15--20 Myr. We also predict 9--12 SNe have occurred in this lifetime. We tried the same approach for the OB association though there was an insufficient number of stars to constrain the IMF.

\subsubsection{Observed versus expected mass/energy comparisons}
We compare our observed SB mass and energy gathered from spectral analysis of the hot X-ray gas to the predicted mass/energy input by the stellar population. A summary of the energy budget is outlined in Table \ref{tab:energy}. The total energy stored in the gas is $(3.56 \pm 0.90)\times 10^{51}$ erg while the predicted injected energy by the clusters is estimated at $(8.72 \pm 3.02)\times 10^{51}$ erg. Hence, the observed energy at the current epoch is almost half of the predicted injected energy from the clusters. 

The stellar clusters are estimated to have injected ($123-184$) M$_\odot$ through SNe and ($233-259$) M$_\odot$ through winds predicted from the output of the \emph{Starburst99} code. Even considering the uncertainty on the observed mass from spectral analysis of ($3800 \pm 970$) M$_\odot$, the mass contribution from the stellar population is much lower.

Assuming that these clusters are the powering sources, there are clearly differences between the observed and expected masses and energies. The energy discrepancy is likely due to the energy losses in the blowouts observed in the region (see Fig. \ref{fig:multi-wavelength}), where the hot X-ray gas has already cooled sufficiently and is not contributing to the total observed energy budget. This mass difference is likely due to mass-loading of the gas by evaporation of entrained interstellar clouds and/or turbulent mixing of material from the cold supershell \citep{weaver1977interstellar}. The injected mass by stellar winds and SNe is much lower compared to that in the expanding HI shell, consistent with HI shells in the Galaxy \citep{2005A&A...437..101E}.

Large discrepancies between the expected input and observed masses/energies are commonplace in the literature. In the N 206 SB \citep{kavanagh2012xmm}, the expected mass input contributes $\sim$40\% of the observed mass with an observed to expected input energy fraction of $\sim$70\%. \cite{cooper2004energy} find that the expected stellar mechanical energy far exceeds the energy stored in the DEM L192 superbubble by a factor of 3. Clearly, we are observing a similar energy discrepancy in the \spbub{} SB. 

It is still unclear whether NGC~1767 and NGC~1782 are the clusters that have driven the evolution of the SB throughout its lifetime. It seems unlikely that  NGC~1782 could given its location on the north eastern extremity of the MSFR. NGC~1767 is a better candidate, located at the northern shell of the SB region. Such off-centre powering clusters are also observed in other SBs such as N~206 \citep{kavanagh2012xmm}, though NGC~1767 is somewhat older and less massive than other cases. As shown in Fig~\ref{fig:bright_stars}, there are also other massive stars scattered throughout the region and the OB association in the western shell. Stellar winds and supernovae from these populations have undoubtedly also contributed to the evolution of this region, though the lack of an age constraint makes it difficult to put their role in context. Given the issues with NGC~1767 as the powering source, we suggest that the morphology of the region was originally shaped by an older central massive stellar population, the remnants of which are now scattered throughout or lost to supernovae, with NGC~1767, NGC~1782, and the OB association now providing the energy input necessary power the diffuse X-ray emission in the region.

\begin{table}[htb]
    \centering
    \caption{Comparisons between observed/expected energies and masses of the \spbub{} SB region.}
    \begin{tabular}{lcc}
    \toprule
     & Energy & Mass \\
     & ($10^{51}$ erg) & (M$_\odot$) \\ \midrule
     
    \multicolumn{3}{c}{Observed} \\ \midrule
    X-ray gas & $3.0\pm 0.8$ & $3800 \pm 970$\\
    H$\alpha$ shell kinetic energy & $0.76\pm 0.07$& - \\
    HI shell kinetic energy & $\lesssim0.1$ & - \\
    Total observed & $3.86\pm 0.9$ & $3800 \pm 970$ \\ \midrule
    
    \multicolumn{3}{c}{Expected} \\ \midrule
    Stellar wind input & $\leq 0.01$ & $233-259$ \\
    SNe input & $12-19$ & $123-184$\\
    Total input & $19.75-26.71$ & $360-446$\\
    \bottomrule
    \end{tabular}
    \tablefoot{The expected inputs are the combination of both NGC~1767 and NGC~1782. The expected energy/mass input is from the output of \emph{Starburst99} simulations.}
    \label{tab:energy}
\end{table}

\subsection{SNR candidates}
There are two known SNR candidates in the MSFR. We now briefly describe these candidates and the implications for their classification using the {\it XMM-Newton} data.

\label{sect:mcsnr_cand}
\subsubsection{MCSNR~J0457-6923}
MCSNR~J0457-6923 is a candidate supernova remnant located in the north-eastern region of the \spbub{} superbubble (see Fig. \ref{fig:multi-wavelength}). It was classified based on its optical shell morphology with enhanced \sii{}/H$\alpha$ ratio $>$ 0.4 and faint 20~cm radio emission \citep{bozzetto2017statistical}. \citet{Zangrandi2024} used eROSITA data to search for the X-ray counterpart to this SNR candidate but could not confirm a counterpart given that it is embedded in the HII region (see their Fig.~E.4~(i)\footnote{The figure is available at \url{https://zenodo.org/records/13951085}}). In spite of the deeper {\it XMM-Newton} observations presented here, we also cannot confirm MCSNR~J0457-6923 as an SNR, finding no clear enhancement of emission in the bounds of the object.

\subsubsection{MCSNR~J0455-6925}
MCSNR~J0455-6925 is a candidate SNR with a bi-lateral morphology embedded in the western shell region (Fig. \ref{fig:multi-wavelength}), identified in \citet{Cotton2026} using the sensitive MeerKAT 1.3~GHz survey of the LMC (see their Fig.~7). The authors were unable to measure a reliable spectral index for the object, though the lack of H$\alpha$ counterpart suggested a non-thermal origin for the radio emission, indicative of an SNR. In addition, comparisons to the {\it XMM-Newton} data in \citet{Cotton2026} and this work showed no clear X-ray counterpart to the candidate. This is unsurprising given its location in the extended western shell emission. 

\section{Summary}
\label{sec:sec5}
We have detected soft diffuse X-ray emission using \emph{XMM-Newton} data of the \spbub{} \ion{H}{II} region. We have analysed this emission in a multi-wavelength context using optical H$\alpha$ data from MCELS, radio data from ATCA, and infrared data from \emph{Spitzer}. We presented the first analysis of the \spbub{} superbubble which can be summarised by the following:

\begin{enumerate}
    \item The X-ray emission displays a complex morphology with a central region mostly surrounded by a H$\alpha$ shell which we label the \spbub{} SB. We found three blowout regions from the SB into the east, west, and possibly south.
    \item We performed spectral analysis of the X-ray emitting gas, including the blowout regions. Our best-fit models confirm a hot shocked gas with a temperature of $kT=0.17~(0.16-0.18)$ keV, a gas density $n_H=0.16 \pm 0.04$ cm$^{-3}$ and a hydrogen column density of $N_H=0.56~(0.50-0.61) \times 10^{22}$ cm$^{-2}$. Additionally, we estimated the pressure and energy of the \spbub{} SB eastern region to be $P/k= (7.45 \pm 1.96) \times 10^5$ cm$^{-3}$ K and $(3.02 \pm 0.79)\times 10^{51}$ erg, respectively with a total mass of $(3800\pm970)$ M$_\odot$.
    \item We compared the energy and mass stored in the X-ray emitting gas with the predictions based on the stellar population. We identified two clusters along the north edge of the region, NGC 1767 and NGC 1782, which have estimated ages of 10 -- 20 Myr with stars of masses 1 - 20 M$_\odot$.  
    Using the stellar population synthesis code \emph{Starburst99} we estimated the energy and mass input of the NGC 1767 and NGC 1782 clusters. We predicted that between 12 and 19 supernovae have occurred in these clusters, which are likely driving the evolution of the region at the current epoch. In total, we estimated that the stellar population has injected $360-446$ M$_\odot$ of mass and $(19.75-26.71)\times 10^{51}$ erg of energy into \spbub{}. This value is much higher than the observed energy, consistent with the energy discrepancy observed in other LMC SBs. 

    \item We could not find the definitive powering source for the SB. NGC~1767 is the most likely candidate, albeit of lower mass and older than powering sources in other SBs. There are also other massive stars dispersed through the region, including the OB association in the western shell. We suggest that the morphology of the region was originally shaped by an older central massive stellar population which is now dispersed through the region or lost to supernovae, and that NGC~1767 and NGC~1782 are now supplying mass and energy.
\end{enumerate} 

\begin{acknowledgements}
     This research is based on data obtained with \emph{XMM-Newton}, an ESA science mission with instruments and contributions directly funded by ESA Member States and NASA. P.J. is funded through the John \& Pat Hume Doctoral scholarships from Maynooth University.
\end{acknowledgements}

\bibliographystyle{aa}
\bibliography{references}

@ARTICLE{Butt2008,
       author = {{Butt}, Yousaf M. and {Bykov}, Andrei M.},
        title = "{A Cosmic-Ray Resolution to the Superbubble Energy Crisis}",
      journal = {\apjl},
         year = 2008,
        month = apr,
       volume = {677},
       number = {1},
        pages = {L21},
          doi = {10.1086/587875},
archivePrefix = {arXiv},
       eprint = {0802.3805},
 primaryClass = {astro-ph},
       adsurl = {https://ui.adsabs.harvard.edu/abs/2008ApJ...677L..21B}
}

@ARTICLE{Cotton2026,
       author = {{Cotton}, W.~D. and {Rajabpour}, N. and {Filipovi{\'c}}, M.~D. and {Camilo}, F. and {Alsaberi}, R.~Z.~E. and {Bester}, L.~H. and {Bradley}, A.~C. and {Crawford}, E.~J. and {Ghavam}, M. and {Khattab}, O.~K. and {Smeaton}, Z.~J. and {Smirnov}, O.~M. and {van Loon}, J. Th and {Velovi{\'c}}, V.},
        title = "{The MeerKAT 1.3 GHz survey of the Large Magellanic Cloud}",
      journal = {\mnras},
         year = 2026,
        month = jul,
       volume = {549},
       number = {4},
          eid = {stag963},
        pages = {stag963},
          doi = {10.1093/mnras/stag963},
archivePrefix = {arXiv},
       eprint = {2602.21353},
 primaryClass = {astro-ph.GA},
       adsurl = {https://ui.adsabs.harvard.edu/abs/2026MNRAS.549ag963C}
}

@ARTICLE{Kim2017,
       author = {{Kim}, Chang-Goo and {Ostriker}, Eve C. and {Raileanu}, Roberta},
        title = "{Superbubbles in the Multiphase ISM and the Loading of Galactic Winds}",
      journal = {\apj},
         year = 2017,
        month = jan,
       volume = {834},
       number = {1},
          eid = {25},
        pages = {25},
          doi = {10.3847/1538-4357/834/1/25},
archivePrefix = {arXiv},
       eprint = {1610.03092},
 primaryClass = {astro-ph.GA},
       adsurl = {https://ui.adsabs.harvard.edu/abs/2017ApJ...834...25K}
}

@ARTICLE{Breit2006,
       author = {{Breitschwerdt}, D. and {de Avillez}, M.~A.},
        title = "{The history and future of the Local and Loop I bubbles}",
      journal = {\aap},
         year = 2006,
        month = jun,
       volume = {452},
       number = {1},
        pages = {L1-L5},
          doi = {10.1051/0004-6361:20064989},
archivePrefix = {arXiv},
       eprint = {astro-ph/0604162},
 primaryClass = {astro-ph},
       adsurl = {https://ui.adsabs.harvard.edu/abs/2006A&A...452L...1B}
}

@ARTICLE{Oey2004,
       author = {{Oey}, M.~S. and {Garc{\'\i}a-Segura}, G.},
        title = "{Ambient Interstellar Pressure and Superbubble Evolution}",
      journal = {\apj},
         year = 2004,
        month = sep,
       volume = {613},
       number = {1},
        pages = {302-311},
          doi = {10.1086/421483},
archivePrefix = {arXiv},
       eprint = {astro-ph/0403713},
 primaryClass = {astro-ph},
       adsurl = {https://ui.adsabs.harvard.edu/abs/2004ApJ...613..302O}
}

@ARTICLE{Oey1996,
       author = {{Oey}, M.~S.},
        title = "{The Dynamics of Superbubbles in the Large Magellanic Cloud}",
      journal = {\apj},
         year = 1996,
        month = aug,
       volume = {467},
        pages = {666},
          doi = {10.1086/177642},
       adsurl = {https://ui.adsabs.harvard.edu/abs/1996ApJ...467..666O}
}

@ARTICLE{Points2024,
       author = {{Points}, Sean D. and {Long}, Knox S. and {Blair}, William P. and {Williams}, Rosa and {Chu}, You-Hua and {Winkler}, P. Frank and {White}, Richard L. and {Rest}, Armin and {Li}, Chuan-Jui and {Valdes}, Francisco},
        title = "{The Dark Energy Camera Magellanic Clouds Emission-line Survey}",
      journal = {\apj},
         year = 2024,
        month = oct,
       volume = {974},
       number = {1},
          eid = {70},
        pages = {70},
          doi = {10.3847/1538-4357/ad6766},
archivePrefix = {arXiv},
       eprint = {2409.04846},
 primaryClass = {astro-ph.GA},
       adsurl = {https://ui.adsabs.harvard.edu/abs/2024ApJ...974...70P}
}

@ARTICLE{Zangrandi2024,
       author = {{Zangrandi}, Federico and {Jurk}, Katharina and {Sasaki}, Manami and {Knies}, Jonathan and {Filipovi{\'c}}, Miroslav D. and {Haberl}, Frank and {Kavanagh}, Patrick and {Maitra}, Chandreyee and {Maggi}, Pierre and {Saeedi}, Sara and {Bernreuther}, Dominic and {Koribalski}, B{\"a}rbel S. and {Points}, Sean and {Staveley-Smith}, Lister},
        title = "{First study of the supernova remnant population in the Large Magellanic Cloud with eROSITA}",
      journal = {\aap},
         year = 2024,
        month = dec,
       volume = {692},
          eid = {A237},
        pages = {A237},
          doi = {10.1051/0004-6361/202348868},
archivePrefix = {arXiv},
       eprint = {2401.17307},
 primaryClass = {astro-ph.HE},
       adsurl = {https://ui.adsabs.harvard.edu/abs/2024A&A...692A.237Z}
}

@phdthesis{maggi2015population,
  title={On the population of supernova remnants in the Large Magellanic Cloud observed with XMM-Newton},
  author={Maggi, Pierre},
  year={2015},
  school={Technische Universit{\"a}t M{\"u}nchen}
}

@ARTICLE{Maggi_LMC,
       author = {{Maggi}, P. and {Haberl}, F. and {Kavanagh}, P.~J. and {Sasaki}, M. and {Bozzetto}, L.~M. and {Filipovi{\'c}}, M.~D. and {Vasilopoulos}, G. and {Pietsch}, W. and {Points}, S.~D. and {Chu}, Y. -H. and {Dickel}, J. and {Ehle}, M. and {Williams}, R. and {Greiner}, J.},
        title = "{The population of X-ray supernova remnants in the Large Magellanic Cloud}",
      journal = {\aap},
         year = 2016,
        month = jan,
       volume = {585},
          eid = {A162},
        pages = {A162},
          doi = {10.1051/0004-6361/201526932},
archivePrefix = {arXiv},
       eprint = {1509.09223},
 primaryClass = {astro-ph.HE},
       adsurl = {https://ui.adsabs.harvard.edu/abs/2016A&A...585A.162M}
}

@software{pyxspec,
       author = {{Gordon}, Craig and {Arnaud}, Keith},
        title = "{PyXspec: Python interface to XSPEC spectral-fitting program}",
 howpublished = {ASCL},
         year = 2021,
        month = jan,
          eid = {ascl:2101.014},
       adsurl = {https://ui.adsabs.harvard.edu/abs/2021ascl.soft01014G}
}

@ARTICLE{wilms,
       author = {{Wilms}, J. and {Allen}, A. and {McCray}, R.},
        title = "{On the Absorption of X-Rays in the Interstellar Medium}",
      journal = {\apj},
         year = 2000,
        month = oct,
       volume = {542},
       number = {2},
        pages = {914-924},
          doi = {10.1086/317016},
archivePrefix = {arXiv},
       eprint = {astro-ph/0008425},
 primaryClass = {astro-ph},
       adsurl = {https://ui.adsabs.harvard.edu/abs/2000ApJ...542..914W}
}

@ARTICLE{balucinska,
       author = {{Balucinska-Church}, Monika and {McCammon}, Dan},
        title = "{Photoelectric Absorption Cross Sections with Variable Abundances}",
      journal = {\apj},
         year = 1992,
        month = dec,
       volume = {400},
        pages = {699},
          doi = {10.1086/172032},
       adsurl = {https://ui.adsabs.harvard.edu/abs/1992ApJ...400..699B}
}

@article{snowden2008catalog,
  title={A catalog of galaxy clusters observed by XMM-Newton},
  author={Snowden, SL and Mushotzky, RF and Kuntz, KD and Davis, David S},
  journal={\aap},
  volume={478},
  number={2},
  pages={615--658},
  year={2008},
  publisher={EDP Sciences}
}

@ARTICLE{agnpow_chen,
       author = {{Chen}, L. -W. and {Fabian}, A.~C. and {Gendreau}, K.~C.},
        title = "{ASCA and ROSAT observations of the QSF3 field: the X-ray background in the 0.1-7 keV band}",
      journal = {\mnras},
         year = 1997,
        month = mar,
       volume = {285},
       number = {3},
        pages = {449-471},
          doi = {10.1093/mnras/285.3.449},
       adsurl = {https://ui.adsabs.harvard.edu/abs/1997MNRAS.285..449C}
}

@article{smith2001collisional,
  title={Collisional plasma models with APEC/APED: emission-line diagnostics of hydrogen-like and helium-like ions},
  author={Smith, Randall K and Brickhouse, Nancy S and Liedahl, Duane A and Raymond, John C},
  journal={\apj},
  volume={556},
  number={2},
  pages={L91},
  year={2001},
  publisher={IOP Publishing}
}

@article{dickey1990hi,
  title={HI in the Galaxy},
  author={Dickey, John M and Lockman, Felix J},
  journal={\araa},
  volume={28},
  number={1},
  pages={215--259},
  year={1990},
  publisher={Annual Reviews 4139 El Camino Way, PO Box 10139, Palo Alto, CA 94303-0139, USA}
}

@article{sasaki2011xmm,
  title={XMM-Newton observations of the superbubble in N 158 in the LMC},
  author={Sasaki, Manami and Breitschwerdt, Dieter and Baumgartner, Verena and Haberl, Frank},
  journal={\aap},
  volume={528},
  pages={A136},
  year={2011},
  publisher={EDP Sciences}
}

@article{pietrzynski2019distance,
  title={A distance to the Large Magellanic Cloud that is precise to one per cent},
  author={Pietrzy{\'n}ski, G and Graczyk, D and Gallenne, A and Gieren, W and Thompson, IB and Pilecki, B and Karczmarek, P and G{\'o}rski, M and Suchomska, K and Taormina, M and others},
  journal={\nat},
  volume={567},
  number={7747},
  pages={200--203},
  year={2019},
  publisher={Nature Publishing Group UK London}
}

@article{kavanagh2012xmm,
  title={XMM-Newton view of the N 206 superbubble in the Large Magellanic Cloud},
  author={Kavanagh, Patrick J and Sasaki, Manami and Points, SD},
  journal={\aap},
  volume={547},
  pages={A19},
  year={2012},
  publisher={EDP Sciences}
}

@article{kim2023tigress,
  title={Introducing TIGRESS-NCR. I. Coregulation of the Multiphase Interstellar Medium and Star Formation Rates},
  author={Kim, Chang-Goo and Kim, Jeong-Gyu and Gong, Munan and Ostriker, Eve C},
  journal={\apj},
  volume={946},
  number={1},
  pages={3},
  year={2023},
  publisher={IOP Publishing}
}

@article{cooper2004energy,
  title={Energy crisis in the superbubble DEM L192 (N51D)},
  author={Cooper, Randall L and Guerrero, Mart{\'\i}n A and Chu, You-Hua and Chen, C-H Rosie and Dunne, Bryan C},
  journal={\apj},
  volume={605},
  number={2},
  pages={751},
  year={2004},
  publisher={IOP Publishing}
}

@article{maddox2009nonthermal,
  title={Nonthermal X-ray emission in the N11 superbubble in the large Magellanic cloud},
  author={Maddox, LA and Williams, RM and Dunne, BC and Chu, Y-H},
  journal={\apj},
  volume={699},
  number={1},
  pages={911},
  year={2009},
  publisher={IOP Publishing}
}

@article{jaskot2011observational,
  title={Observational Constraints on Superbubble X-ray Energy Budgets},
  author={Jaskot, AE and Strickland, DK and Oey, MS and Chu, Y-H and Garc{\'\i}a-Segura, G},
  journal={\apj},
  volume={729},
  number={1},
  pages={28},
  year={2011},
  publisher={IOP Publishing}
}

@article{dunne2001x,
  title={X-ray Emission from Superbubbles in the LMC: VI. A Sample of 13 Superbubbles},
  author={Dunne, BC and Points, SD and Chu, Yo-H},
  journal={ApJS},
  volume={136},
  pages={119--135},
  year={2001}
}

@article{werner2004spitzer,
  title={The Spitzer space telescope mission},
  author={Werner, Michael W and Roellig, Thomas L and Low, FJ and Rieke, George H and Rieke, M and Hoffmann, WF and Young, E and Houck, JR and Brandl, B and Fazio, GG and others},
  journal={\apjs},
  volume={154},
  number={1},
  pages={1},
  year={2004},
  publisher={IOP Publishing}
}

@article{fazio2004irac,
  title={The infrared array camera (IRAC) for the spitzer space telescope},
  author={Fazio, GG and Hora, JL and Allen, LE and Ashby, MLN and Barmby, P and Deutsch, LK and Huang, J-S and Kleiner, S and Marengo, M and Megeath, ST and others},
  journal={\apjs},
  volume={154},
  number={1},
  pages={10},
  year={2004},
  publisher={IOP Publishing}
}

@article{rieke2004mips,
  title={The multiband imaging photometer for Spitzer (MIPS)},
  author={Rieke, GH and Young, ET and Engelbracht, CW and Kelly, DM and Low, FJ and Haller, EE and Beeman, JW and Gordon, KD and Stansberry, JA and Misselt, KA and others},
  journal={\apjs},
  volume={154},
  number={1},
  pages={25},
  year={2004},
  publisher={IOP Publishing}
}

@article{weaver1977interstellar,
  title={Interstellar bubbles. II-Structure and evolution},
  author={Weaver, Robert and McCray, Richard and Castor, John and Shapiro, Paul and Moore, Robert},
  journal={\apj},
  volume={218},
  pages={377--395},
  year={1977}
}

@article{townsley2011integrated,
  title={The integrated diffuse X-ray emission of the Carina Nebula compared to other massive star-forming regions},
  author={Townsley, Leisa K and Broos, Patrick S and Chu, You-Hua and Gruendl, Robert A and Oey, Md S and Pittard, Julian M},
  journal={\apjs},
  volume={194},
  number={1},
  pages={16},
  year={2011},
  publisher={IOP Publishing}
}

@article{wang1991x,
  title={An X-ray image of the violent interstellar medium in 30 Doradus},
  author={Wang, Q and Helfand, David J},
  journal={\apj},
  volume={370},
  pages={541--550},
  year={1991}
}

@article{townsley2011chandra,
  title={The Chandra Carina Complex Project: Deciphering the enigma of Carina's diffuse X-ray emission},
  author={Townsley, Leisa K and Broos, Patrick S and Chu, You-Hua and Gagn{\'e}, Marc and Garmire, Gordon P and Gruendl, Robert A and Hamaguchi, Kenji and Mac Low, Mordecai-Mark and Montmerle, Thierry and Naz{\'e}, Ya{\"e}l and others},
  journal={\apjs},
  volume={194},
  number={1},
  pages={15},
  year={2011},
  publisher={IOP Publishing}
}

@ARTICLE{oey1998dem192,
       author = {{Oey}, M.~S. and {Smedley}, Shona A.},
        title = "{Shell Formation and Star Formation in Superbubble DEM 192}",
      journal = {\aj},
         year = 1998,
        month = sep,
       volume = {116},
       number = {3},
        pages = {1263-1274},
          doi = {10.1086/300498},
archivePrefix = {arXiv},
       eprint = {astro-ph/9806009},
 primaryClass = {astro-ph},
       adsurl = {https://ui.adsabs.harvard.edu/abs/1998AJ....116.1263O}
}

@ARTICLE{reyesN70N185,
       author = {{Reyes-Iturbide}, J. and {Rosado}, M. and {Rodr{\'\i}guez-Gonz{\'a}lez}, A. and {Vel{\'a}zquez}, P.~F. and {S{\'a}nchez-Cruces}, M. and {Ambrocio-Cruz}, P.},
        title = "{Diffuse X-Ray Emission from the Superbubbles N70 and N185 in the Large Magellanic Cloud}",
      journal = {\aj},
         year = 2014,
        month = nov,
       volume = {148},
       number = {5},
          eid = {102},
        pages = {102},
          doi = {10.1088/0004-6256/148/5/102},
archivePrefix = {arXiv},
       eprint = {1407.7586},
 primaryClass = {astro-ph.HE},
       adsurl = {https://ui.adsabs.harvard.edu/abs/2014AJ....148..102R}
}

@article{kumar2008ccd,
  title={CCD photometric and mass function study of nine young Large Magellanic Cloud star clusters},
  author={Kumar, B and Sagar, R and Melnick, J},
  journal={\mnras},
  volume={386},
  number={3},
  pages={1380--1397},
  year={2008},
  publisher={Blackwell Publishing Ltd Oxford, UK}
}

@article{zaritsky2004magellanic,
  title={The Magellanic clouds photometric survey: the Large Magellanic Cloud stellar catalog and extinction map},
  author={Zaritsky, Dennis and Harris, Jason and Thompson, Ian B and Grebel, Eva K},
  journal={\aj},
  volume={128},
  number={4},
  pages={1606},
  year={2004},
  publisher={IOP Publishing}
}

@ARTICLE{mist0,
       author = {{Dotter}, Aaron},
        title = "{MESA Isochrones and Stellar Tracks (MIST) 0: Methods for the Construction of Stellar Isochrones}",
      journal = {\apjs},
         year = 2016,
        month = jan,
       volume = {222},
       number = {1},
          eid = {8},
        pages = {8},
          doi = {10.3847/0067-0049/222/1/8},
archivePrefix = {arXiv},
       eprint = {1601.05144},
 primaryClass = {astro-ph.SR},
       adsurl = {https://ui.adsabs.harvard.edu/abs/2016ApJS..222....8D}
}

@ARTICLE{mist1,
       author = {{Choi}, Jieun and {Dotter}, Aaron and {Conroy}, Charlie and {Cantiello}, Matteo and {Paxton}, Bill and {Johnson}, Benjamin D.},
        title = "{Mesa Isochrones and Stellar Tracks (MIST). I. Solar-scaled Models}",
      journal = {\apj},
         year = 2016,
        month = jun,
       volume = {823},
       number = {2},
          eid = {102},
        pages = {102},
          doi = {10.3847/0004-637X/823/2/102},
archivePrefix = {arXiv},
       eprint = {1604.08592},
 primaryClass = {astro-ph.SR},
       adsurl = {https://ui.adsabs.harvard.edu/abs/2016ApJ...823..102C}
}

@article{choudhury2016photometric,
  title={Photometric metallicity map of the Large Magellanic Cloud},
  author={Choudhury, Samyaday and Subramaniam, Annapurni and Cole, Andrew A},
  journal={\mnras},
  volume={455},
  number={2},
  pages={1855--1880},
  year={2016},
  publisher={Oxford University Press}
}

@ARTICLE{furuta2022extinction,
       author = {{Furuta}, Takuya and {Kaneda}, Hidehiro and {Kokusho}, Takuma and {Nakajima}, Yasushi and {Fukui}, Yasuo and {Tsuge}, Kisetsu},
        title = "{Three-dimensional geometry and dust/gas ratios in massive star-forming regions over the entire LMC as revealed by the IRSF/SIRIUS survey}",
      journal = {\pasj},
         year = 2022,
        month = jun,
       volume = {74},
       number = {3},
        pages = {639-655},
          doi = {10.1093/pasj/psac025},
archivePrefix = {arXiv},
       eprint = {2203.03206},
 primaryClass = {astro-ph.GA},
       adsurl = {https://ui.adsabs.harvard.edu/abs/2022PASJ...74..639F}
}

@ARTICLE{imara2007extinction,
       author = {{Imara}, Nia and {Blitz}, Leo},
        title = "{Extinction in the Large Magellanic Cloud}",
      journal = {\apj},
         year = 2007,
        month = jun,
       volume = {662},
       number = {2},
        pages = {969-979},
          doi = {10.1086/517911},
archivePrefix = {arXiv},
       eprint = {astro-ph/0703421},
 primaryClass = {astro-ph},
       adsurl = {https://ui.adsabs.harvard.edu/abs/2007ApJ...662..969I}
}

@ARTICLE{chabrier2003,
       author = {{Chabrier}, Gilles},
        title = "{Galactic Stellar and Substellar Initial Mass Function}",
      journal = {\pasp},
         year = 2003,
        month = jul,
       volume = {115},
       number = {809},
        pages = {763-795},
          doi = {10.1086/376392},
archivePrefix = {arXiv},
       eprint = {astro-ph/0304382},
 primaryClass = {astro-ph},
       adsurl = {https://ui.adsabs.harvard.edu/abs/2003PASP..115..763C}
}

@ARTICLE{chu1990model,
       author = {{Chu}, You-Hua and {Mac Low}, Mordecai-Mark},
        title = "{X-Rays from Superbubbles in the Large Magellanic Cloud}",
      journal = {\apj},
         year = 1990,
        month = dec,
       volume = {365},
        pages = {510},
          doi = {10.1086/169505},
       adsurl = {https://ui.adsabs.harvard.edu/abs/1990ApJ...365..510C}
}

@article{nikolaev2004geometry,
  title={Geometry of the large magellanic cloud disk: Results from macho and the two micron all sky survey},
  author={Nikolaev, S and Drake, AJ and Keller, SC and Cook, Kem H and Dalal, N and Griest, K and Welch, DL and Kanbur, SM},
  journal={\apj},
  volume={601},
  number={1},
  pages={260},
  year={2004},
  publisher={IOP Publishing}
}

@ARTICLE{haberl2022binaries,
       author = {{Haberl}, F. and {Maitra}, C. and {Vasilopoulos}, G. and {Maggi}, P. and {Udalski}, A. and {Monageng}, I.~M. and {Buckley}, D.~A.~H.},
        title = "{Three new high-mass X-ray binaries in the Large Magellanic Cloud}",
      journal = {\aap},
         year = 2022,
        month = jun,
       volume = {662},
          eid = {A22},
        pages = {A22},
          doi = {10.1051/0004-6361/202243301},
archivePrefix = {arXiv},
       eprint = {2203.00625},
 primaryClass = {astro-ph.SR},
       adsurl = {https://ui.adsabs.harvard.edu/abs/2022A&A...662A..22H}
}

@ARTICLE{meaburn1980shells,
       author = {{Meaburn}, J.},
        title = "{The giant and supergiant shells of the Magellanic Clouds.}",
      journal = {\mnras},
         year = 1980,
        month = aug,
       volume = {192},
        pages = {365-375},
          doi = {10.1093/mnras/192.3.365},
       adsurl = {https://ui.adsabs.harvard.edu/abs/1980MNRAS.192..365M}
}

@ARTICLE{sage2006,
       author = {{Meixner}, Margaret and {Gordon}, Karl D. and {Indebetouw}, Remy and {Hora}, Joseph L. and {Whitney}, Barbara and {Blum}, Robert and {Reach}, William and {Bernard}, Jean-Philippe and {Meade}, Marilyn and {Babler}, Brian and {Engelbracht}, Charles W. and {For}, Bi-Qing and {Misselt}, Karl and {Vijh}, Uma and {Leitherer}, Claus and {Cohen}, Martin and {Churchwell}, Ed B. and {Boulanger}, Francois and {Frogel}, Jay A. and {Fukui}, Yasuo and {Gallagher}, Jay and {Gorjian}, Varoujan and {Harris}, Jason and {Kelly}, Douglas and {Kawamura}, Akiko and {Kim}, SoYoung and {Latter}, William B. and {Madden}, Suzanne and {Markwick-Kemper}, Ciska and {Mizuno}, Akira and {Mizuno}, Norikazu and {Mould}, Jeremy and {Nota}, Antonella and {Oey}, M.~S. and {Olsen}, Knut and {Onishi}, Toshikazu and {Paladini}, Roberta and {Panagia}, Nino and {Perez-Gonzalez}, Pablo and {Shibai}, Hiroshi and {Sato}, Shuji and {Smith}, Linda and {Staveley-Smith}, Lister and {Tielens}, A.~G.~G.~M. and {Ueta}, Toshiya and {van Dyk}, Schuyler and {Volk}, Kevin and {Werner}, Michael and {Zaritsky}, Dennis},
        title = "{Spitzer Survey of the Large Magellanic Cloud: Surveying the Agents of a Galaxy's Evolution (SAGE). I. Overview and Initial Results}",
      journal = {\aj},
         year = 2006,
        month = dec,
       volume = {132},
       number = {6},
        pages = {2268-2288},
          doi = {10.1086/508185},
archivePrefix = {arXiv},
       eprint = {astro-ph/0606356},
 primaryClass = {astro-ph},
       adsurl = {https://ui.adsabs.harvard.edu/abs/2006AJ....132.2268M}
}

@ARTICLE{maclow1988superbubbles,
       author = {{Mac Low}, Mordecai-Mark and {McCray}, Richard},
        title = "{Superbubbles in Disk Galaxies}",
      journal = {\apj},
         year = 1988,
        month = jan,
       volume = {324},
        pages = {776},
          doi = {10.1086/165936},
       adsurl = {https://ui.adsabs.harvard.edu/abs/1988ApJ...324..776M}
}

@ARTICLE{kavanagh2020review,
       author = {{Kavanagh}, Patrick J.},
        title = "{Thermal and non-thermal X-ray emission from stellar clusters and superbubbles}",
      journal = {\apss},
         year = 2020,
        month = jan,
       volume = {365},
       number = {1},
          eid = {6},
        pages = {6},
          doi = {10.1007/s10509-019-3719-5},
       adsurl = {https://ui.adsabs.harvard.edu/abs/2020Ap&SS.365....6K}
}

@ARTICLE{leitherer2014starburst,
       author = {{Leitherer}, Claus and {Ekstr{\"o}m}, Sylvia and {Meynet}, Georges and {Schaerer}, Daniel and {Agienko}, Katerina B. and {Levesque}, Emily M.},
        title = "{The Effects of Stellar Rotation. II. A Comprehensive Set of Starburst99 Models}",
      journal = {\apjs},
         year = 2014,
        month = may,
       volume = {212},
       number = {1},
          eid = {14},
        pages = {14},
          doi = {10.1088/0067-0049/212/1/14},
archivePrefix = {arXiv},
       eprint = {1403.5444},
 primaryClass = {astro-ph.GA},
       adsurl = {https://ui.adsabs.harvard.edu/abs/2014ApJS..212...14L}
}

@ARTICLE{leitherer1999starburst,
       author = {{Leitherer}, Claus and {Schaerer}, Daniel and {Goldader}, Jeffrey D. and {Delgado}, Rosa M. Gonz{\'a}lez and {Robert}, Carmelle and {Kune}, Denis Foo and {de Mello}, Du{\'\i}lia F. and {Devost}, Daniel and {Heckman}, Timothy M.},
        title = "{Starburst99: Synthesis Models for Galaxies with Active Star Formation}",
      journal = {\apjs},
         year = 1999,
        month = jul,
       volume = {123},
       number = {1},
        pages = {3-40},
          doi = {10.1086/313233},
archivePrefix = {arXiv},
       eprint = {astro-ph/9902334},
 primaryClass = {astro-ph},
       adsurl = {https://ui.adsabs.harvard.edu/abs/1999ApJS..123....3L}
}

@software{larry_bradley_2024_10967176,
  author       = {Larry Bradley and
                  Brigitta Sip{\H o}cz and
                  Thomas Robitaille and
                  Erik Tollerud and
                  Z\`e Vin{\'{\i}}cius and
                  Christoph Deil and
                  Kyle Barbary and
                  Tom J Wilson and
                  Ivo Busko and
                  Axel Donath and
                  Hans Moritz G{\"u}nther and
                  Mihai Cara and
                  P. L. Lim and
                  Sebastian Me{\ss}linger and
                  Zach Burnett and
                  Simon Conseil and
                  Michael Droettboom and
                  Azalee Bostroem and
                  E. M. Bray and
                  Lars Andersen Bratholm and
                  William Jamieson and
                  Adam Ginsburg and
                  Geert Barentsen and
                  Matt Craig and
                  Sergio Pascual and
                  Shivangee Rathi and
                  Marshall Perrin and
                  Brett M. Morris and
                  Gabriel Perren},
  title        = {astropy/photutils: 1.12.0},
  month        = apr,
  year         = 2024,
  publisher    = {Zenodo},
  version      = {1.12.0},
  doi          = {10.5281/zenodo.10967176},
  url          = {https://doi.org/10.5281/zenodo.10967176}
}

@ARTICLE{Frater1992ATCA,
       author = {{Frater}, Robert H. and {Brooks}, John W. and {Whiteoak}, John B.},
        title = "{The Australia Telescope - Overview}",
      journal = {JEEEA},
         year = 1992,
        month = jun,
       volume = {12},
       number = {2},
        pages = {103-112},
       adsurl = {https://ui.adsabs.harvard.edu/abs/1992JEEEA..12..103F}
}

@article{dunne2008m17,
author = {Dunne, Bryan and Chu, You-Hua and Chen, C.-H and Lowry, Justin and Townsley, Leisa and Gruendl, Robert and Guerrero, Martín and Rosado, Margarita},
year = {2008},
month = {12},
pages = {306},
title = {Diffuse X-Ray Emission from the Quiescent Superbubble M17, the Omega Nebula},
volume = {590},
journal = {\apj},
doi = {10.1086/375010}
}

@article{suarez2024galacticsbs3d,
doi = {10.1088/1742-6596/2796/1/012004},
url = {https://doi.org/10.1088/1742-6596/2796/1/012004},
year = {2024},
month = {jul},
publisher = {IOP Publishing},
volume = {2796},
number = {1},
pages = {012004},
author = {Suárez-López, O. and Villares, A. S. and Banda-Barragán, W. E.},
title = {Galactic Superbubbles in 3D: Wind Formation and Cloud Shielding},
journal = {JPCS}
}

@ARTICLE{taoli2024ngc3079,
       author = {{Li}, Jiang-Tao and {Sun}, Wei and {Ji}, Li and {Yang}, Yang},
        title = "{Pressure Balance and Energy Budget of the Nuclear Superbubble of NGC 3079}",
      journal = {\apj},
         year = 2024,
        month = may,
       volume = {966},
       number = {2},
          eid = {239},
        pages = {239},
          doi = {10.3847/1538-4357/ad3af2},
archivePrefix = {arXiv},
       eprint = {2404.03879},
 primaryClass = {astro-ph.GA},
       adsurl = {https://ui.adsabs.harvard.edu/abs/2024ApJ...966..239L}
}

@article{chi2024thermal,
  title={Thermal X-ray emission in the western half of the LMC superbubble 30 Dor C},
  author={Chi, Yi-Heng and Chen, Han-Xiao and Chen, Yang and Meng, Yi-Fan and Zhou, Ping and Sun, Lei and Sun, Wei},
  journal={\mnras},
  volume={530},
  number={4},
  pages={4219--4233},
  year={2024},
  publisher={Oxford University Press}
}

@article{sasaki2022first,
  title={First studies of the diffuse X-ray emission in the Large Magellanic Cloud with eROSITA},
  author={Sasaki, Manami and Knies, Jonathan and Haberl, Frank and Maitra, Chandreyee and Kerp, J{\"u}rgen and Bykov, Andrei M and Dennerl, Konrad and Filipovi{\'c}, Miroslav D and Freyberg, Michael and Koribalski, B{\"a}rbel S and others},
  journal={\aap},
  volume={661},
  pages={A37},
  year={2022},
  publisher={EDP Sciences}
}

@article{bamba2004thermal,
  title={Thermal and nonthermal X-rays from the large magellanic cloud superbubble 30 Doradus C},
  author={Bamba, Aya and Ueno, Masaru and Nakajima, Hiroshi and Koyama, Katsuji},
  journal={\apj},
  volume={602},
  number={1},
  pages={257},
  year={2004},
  publisher={IOP Publishing}
}

@article{bozzetto2017statistical,
  title={Statistical analysis of supernova remnants in the Large Magellanic Cloud},
  author={Bozzetto, Luke M and Filipovi{\'c}, Miroslav D and Vukoti{\'c}, Branislav and Pavlovi{\'c}, Marko Z and Uro{\v{s}}evi{\'c}, Dejan and Kavanagh, Patrick J and Arbutina, Bojan and Maggi, Pierre and Sasaki, Manami and Haberl, Frank and others},
  journal={\apjs},
  volume={230},
  number={1},
  pages={2},
  year={2017},
  publisher={IOP Publishing}
}

@INPROCEEDINGS{mcels1999-smith,
       author = {{Smith}, R.~C. and {MCELS Team}},
        title = "{The UM/CTIO Magellanic Cloud Emission-line Survey}",
    booktitle = {New Views of the Magellanic Clouds},
         year = 1999,
       editor = {{Chu}, Y.-H. and {Suntzeff}, N. and {Hesser}, J. and {Bohlender}, D.},
       series = {IAU Symposium},
       volume = {190},
        month = jan,
        pages = {28},
       adsurl = {https://ui.adsabs.harvard.edu/abs/1999IAUS..190...28S}
}

@ARTICLE{breysacher_1999_catalogue,
       author = {{Breysacher}, J. and {Azzopardi}, M. and {Testor}, G.},
        title = "{The fourth catalogue of Population I Wolf-Rayet stars in the Large Magellanic Cloud}",
      journal = {\aaps},
         year = 1999,
        month = may,
       volume = {137},
        pages = {117-145},
          doi = {10.1051/aas:1999240},
       adsurl = {https://ui.adsabs.harvard.edu/abs/1999A&AS..137..117B}
}

@ARTICLE{henize_1956_deml36,
       author = {{Henize}, Karl G.},
        title = "{Catalogues of H{\ensuremath{\alpha}}-emission Stars and Nebulae in the Magellanic Clouds.}",
      journal = {\apjs},
         year = 1956,
        month = sep,
       volume = {2},
        pages = {315},
          doi = {10.1086/190025},
       adsurl = {https://ui.adsabs.harvard.edu/abs/1956ApJS....2..315H}
}

@ARTICLE{kim_1998_HILMCpeakmap,
       author = {{Kim}, Sungeun and {Staveley-Smith}, Lister and {Dopita}, Michael A. and {Freeman}, Ken C. and {Sault}, Robert J. and {Kesteven}, Mike J. and {McConnell}, David},
        title = "{An H I Aperture Synthesis Mosaic of the Large Magellanic Cloud}",
      journal = {\apj},
         year = 1998,
        month = aug,
       volume = {503},
       number = {2},
        pages = {674-688},
          doi = {10.1086/306030},
       adsurl = {https://ui.adsabs.harvard.edu/abs/1998ApJ...503..674K}
}

@ARTICLE{2005A&A...437..101E,
       author = {{Ehlerov{\'a}}, S. and {Palou{\v{s}}}, J.},
        title = "{H I shells in the outer Milky Way}",
      journal = {\aap},
         year = 2005,
        month = jul,
       volume = {437},
       number = {1},
        pages = {101-112},
          doi = {10.1051/0004-6361:20034389},
archivePrefix = {arXiv},
       eprint = {astro-ph/0503443},
 primaryClass = {astro-ph},
       adsurl = {https://ui.adsabs.harvard.edu/abs/2005A&A...437..101E}
}

\end{document}